\documentclass[11pt]{JHEP3}
\input epsf.tex
\epsfclipon

\usepackage{color}
\let\normalcolor\relax
\usepackage{epsfig}
\usepackage{epstopdf}
\usepackage{epsf}
\input{epsf.sty}
\usepackage{graphicx,amsmath,amssymb}
\usepackage{epsfig,multicol}
\usepackage{multirow}
\allowdisplaybreaks

\usepackage{cite}

\usepackage{bbm,bm,amsmath,amssymb}
 
\usepackage[normalem]{ulem}

\usepackage{comment}
\def\be{\begin{equation}}
\def\ee{\end{equation}}
\def\figs/B{B}
\def\bea{\begin{eqnarray}}
\def\eea{\end{eqnarray}}
\def\bg{\begin{eqnarray}}
\def\nd{\end{eqnarray}}

\def\ln{{\rm log}}

\title{Crisis on Infinite Earths: Short-lived de Sitter Vacua in the String Theory Landscape}
\author{Heliudson Bernardo$^{1}$, Suddhasattwa Brahma$^{1}$, Keshav Dasgupta$^{1}$, Radu 
Tatar$^2$\\
	\vskip.03in
	${}^1$ Department of Physics, McGill University, Montr\'{e}al, Qu\'{e}bec, H3A 2T8, Canada \\
	${}^2$ Department of Mathematical Sciences, University of Liverpool, Liverpool, L69 7ZL, United Kingdom\\	
		{\tt heliudson@hep.physics.mcgill.ca, suddhasattwa.brahma@mcgill.ca}
	~~{\tt  keshav@hep.physics.mcgill.ca, Radu.Tatar@Liverpool.ac.uk}}

\date{\today}
\maketitle

\abstract{We construct {\it purely non-perturbative} anti-de Sitter vacua in string theory which, on uplifting to a de Sitter (dS) one, have a decay time many orders of magnitude smaller than those of standard constructions, such as the KKLT and LVS scenarios. By virtue of being constructed purely from non-perturbative terms, these vacua avoids certain obstructions plaguing other constructions of dS in string theory. This results in a new class of phenomenological dS vacua in string theory with novel distinctive characteristics such as having two maxima. After examining whether these uplifted dS vacua obey the TCC, we revisit some old problems of realization of dS space as a {\it vacuum}. We find that not only is it phenomenologically hard to construct TCC-compatible vacua, but also inherent temporal dependences of the degrees of freedom generically arise in such constructions, amongst other issues. This reinforces the idea that dS, if it exists in string theory, should be a Glauber-Sudarshan state and not a vacuum.}

\begin{document}

\section{Introduction and summary of the results}
The existence of a de Sitter (dS) background in string theory has been a topic of intense research in the last years \cite{Moritz:2017xto,Gautason:2018gln,Hamada:2018qef,Gautason:2019jwq,Hamada:2019ack,Kachru:2019dvo, desitter2, desitter3}. On one hand, having a four dimensional compactification with a positive cosmological constant seems to be mandatory if one would like to connect string theory with the actual phase of accelerated expansion of the universe and with cosmological inflation. On the other hand, a positive cosmological constant is not naturally obtained using the present available tools and constructing a dS solution with total control, over many approximations, turns out to be very difficult \cite{Danielsson:2018ztv}, to put it mildly. 

Apart from the aforementioned approaches starting from a ten dimensional calculation to study the ingredients for constructing a dS solution in four dimensions, models based in the KKLT \cite{KKLT} and Large Volume Scenario (LVS) \cite{Balasubramanian:2005zx} paradigms of moduli stabilization in the GKP framework of type IIB compactification \cite{GKP} were recently subjected to scrutiny in light of the web of swampland conjectures, that aim to classify effective theories with a UV completion into string theory \cite{Ooguri:2006in,Obied:2018sgi,Garg:2018reu,Ooguri:2018wrx}. There are now many different connected conjectures, and some explicit examples of how they are satisfied from a top-down approach \cite{Palti:2019pca, Brennan:2017rbf}. In particular, the refined dS conjecture renders a stable dS solution for the effective field theory basically impossible.  

More recently, based on early works on trans-Plackian issues in a dS inflationary cosmology \cite{Martin:2000xs}, the trans-Planckian censorship conjecture (TCC) \cite{Bedroya:2019tba} was proposed as a swampland criterion \cite{TCC1} and was shown to be connected with other swampland conjectures \cite{Brahma:2019vpl} (see also \cite{Bernardo:2019bbi} for a relation with holography). The TCC states that an EFT can have meta-stable dS vacua as long as their lifetimes obey the bound \cite{TCC1}:
\begin{eqnarray}\label{TCC_lifetime}
	T \le H^{-1} \ln\left(H^{-1}\right)\,,
\end{eqnarray}
and so it is weaker than the dS conjecture\footnote{It is interesting to note that the absence of meta-stable dS vacua is related to the condition for no eternal inflation, as discussed in \cite{Rudelius:2019cfh} (see also \cite{Brahma:2019iyy, Wang:2019eym}) and recently explored in connection to the TCC in \cite{Bedroya:2020rac}.}. Using the Friedmann equation, we can relate the $H$ (the Hubble parameter) to the value of the potential at the minimum. It has been shown in \cite{TCC1} that the behaviour of both the KKLT and LVS potentials are, \`a priori, completely consistent with the TCC at large field values. In other words, the leading behaviour of the potential for these solutions, at large $\varphi$, is of the form $V \sim e^{-\alpha \varphi}$ due to Dine-Seiberg runaway,  as is required by the TCC. However, the lifetime of both these vacua are of the order:
\begin{eqnarray}\label{KKLT_lifetime}
	T \sim \exp\left(c/V_0\right)\,,
\end{eqnarray}
where $c \sim 24\pi^2$. Even considering this $c$ to be an $\mathcal{O}\left(1\right)$ number, this is severely in conflict with the bound \eqref{TCC_lifetime} and thus these stringy solutions are ruled out by the TCC. Based on these findings, it was concluded in  \cite{TCC1} that, assuming the TCC to be correct, either these solutions are not realizable in string theory \textit{or} we need to find new channels for decay of these meta-stable dS solutions. Since the decay of these vacua are considered via the usual Coleman-de Luccia mechanism, it seems unlikely that there would be other decay channels which dominate over these gravitational instantons. 

It is also well-known \cite{Westphal:2007xd} that the lifetime of a wide range of dS solutions in string theory, which include both the KKLT and the LVS case, is exponentially long as given in \eqref{KKLT_lifetime}. The class of models which were considered have a single dS vacuum, after ``uplifting'' an AdS extremum, separated from the Minkowski vacuum at infinite volume. In other words, this lifetime calculation assumes that the meta-stable dS decays via Coleman-de Lucca instanton under the thin-wall approximation, which is justified given the observed small value of the cosmological constant. Thus, the lifetime calculated in this manner is quite general for such potentials, which are then all ruled out by the TCC, as a consequence.

In this work, we attempt to find a potential with a new decay channel such that we have a dS solution with a very short lifetime as required by TCC\footnote{For an alternative proposal, see \cite{Farakos:2020wfc}.}. In section \ref{sec2} we approach the problem from the four dimensional effective theory perspective and, concentrating on the validity of the TCC criterion, the moduli is stabilised via gaugino condensation and standard uplift from AdS to dS is performed. However, as we shall show later in section \ref{sec3}, although we manage to find dS vacua which have extremely small lifetimes as compared to \eqref{KKLT_lifetime}, they still necessarily violate the TCC constraint. Having arrived at the conclusion that it is phenomenologically difficult to satisfy the TCC bound for the dS vacua constructed, we then go on to examine some of the implicit assumptions associated with this construction from an M-theory point of view.

In contrast to the class of models discussed in \cite{Westphal:2007xd}, we shall consider a model in which there would be two AdS minima, prior to uplifting, utilizing the racetrack mechanism. The KL model introduced in \cite{Kallosh:2004yh} has exactly such a behaviour, in which two independent non-perturbative contributions to the superpotential are tuned such that there is a Minkowski and an AdS minimum, and the parameters are then tuned to build a model with two AdS minima (see also \cite{Kallosh:2019axr}). After uplifting, the shallower of the two AdS vacua is converted into a dS one while the other remains AdS. Therefore, the main decay channel for this dS minimum would be towards the other AdS vacuum and this is why we are able to achieve lifetimes much smaller than what is given in \eqref{KKLT_lifetime}. 

Interestingly, as explained in \cite{BlancoPillado:2005fn}, the KL model has some advantages over the KKLT formulation concerning the stability of the potential and the supersymmetry (susy) breaking after the uplift (which is due to the possibility of having a susy Minkowski vacuum prior to the uplift). On the other hand, the minima contructed in the KL model, just as in KKLT, require a non-zero, highly-tuned value of the Gukov-Witten-Vafa superpotential $W_{\text{GVW}}$ generated by the fluxes\footnote{We define 
$G_3 \equiv {\bf F}_3 - {i\over g_b} ~{\bf H}_3$ for vanishing axion and constant type IIB coupling $g_b$. The RR and NS three-form fluxes are denoted by ${\bf F}_3$ and ${\bf H}_3$ respectively.} \cite{GVW, DRS, GKP}:
\begin{equation}
    W_{\text{GVW}} = \int G_{3} \wedge \Omega\,,
\end{equation}
as reviewed in more detail in section \ref{sec2}. This leads to a serious problem as discussed in \cite{Sethi:2017phn}, where it was argued that consistency with the M-theory equations of motion would imply that the fluxes themselves break susy even before the uplift. Denoting $W_0$ as the value of the $W_{\text{GVW}}$ superpotential evaluated at the stabilized values of the complex structure fields, following the arguments of \cite{Sethi:2017phn} there is no reason to expect a perturbative solution to the $10$-d equations of motion when $W_0\neq 0$. From the $4$-d effective theory point of view, we could avoid such obstructions by simply considering the case $W_0 = 0$ that corresponds to starting with a susy-preserving Minkowski vacuum, which at first sight would prevent having two AdS minima as in the KL model.

However, as discussed in next section, restricting ourselves to the $4$-d effective field theory, we propose a way around to still get a new channel for dS decay while having $W_0 = 0$: one can add an extra non-perturbative contribution for the superpotential as compared with the KL model. This modification implies that we would be constructing a purely non-perturbative AdS vacua starting from $4$-d susy Minskowski vacuum. The resulting $4$-d potential can be tuned to have two susy AdS vacua or one Minkowski and one AdS vacua but has two maxima and approches zero from positive values -- a distinct feature  characteristic of this model. After uplifting one of the minima to dS, we shall see both analytically and numerically that the region of parameter space of this model, open to tuning under the starting assumptions, is not compatible with TCC. Thus, even assuming the uplifting procedure to be correct, the final dS does not satisfy the TCC.

From the higher dimensional perspective, given the interplay between gaugino condensation with fluxes and the backreaction of the uplift by anti-branes (see e.g. \cite{Bena:2019mte, Moritz:2017xto}), a proper $10$-d or M-theory analysis is required to check the consistency of the $4$-d solution. To that end, in section \ref{sec4} we step back from the $4$-d effective description and study the possible uplift from a given AdS solution to a dS solution in type IIB. We work within the framework developed in \cite{desitter2, desitter3}, which includes basically all possible higher derivative corrections to the M-theory equations of motion. We find no obstructions to having an AdS vacuum with $W_0=0$, therefore solving the issues raised in \cite{Sethi:2017phn} and allowing our model (before the uplift) to be embedded into M-theory. Unfortunately, we also find that consistency with the M-theory equations of motion, and existence of a hierarchy among  higher order corrections, requires the fluxes to develop a time dependence after the uplift. This seems to indicate that we cannot describe an uplifted dS as a solution of the effective theory, supporting the conclusions in \cite{desitter3} and motivating the description of dS as a Glauber-Sudarshan state proposed in \cite{coherbeta}. The time-dependence of the fluxes in fact appears from the M-theory anomaly cancellation condition:
\bg\label{anomaly}
b_2 \int {\bf G}_4 \wedge {\bf G}_4 + n_b - \bar{n}_b = -b_3 \int {\bf X}_8, \nd
where ${\bf G}_4$ is the four-form G-flux in M-theory; $b_i$ are constant (${\rm M}_p$ dependent\footnote{We use ${\rm M}_p$ to denote the eleven-dimensional Plank mass whereas $M_{\rm Pl}$ is reserved for the four-dimensional Planck mass.}) coefficients defined in \eqref{weisz2}; $(n_b, \bar{n}_b)$ are the number of 
M2 and $\overline{\rm M2}$ -branes;  and ${\bf X}_8$ is a polynomial constructed using quartic order in curvature form as defined in \eqref{polameys}. This form is generically time-{\it dependent} as shown in 
section \ref{X8}, and because of that, the four-form fluxes become time-dependent (there are more subtleties to this which will be explained in section \ref{flox}). Consequently the dual IIB fluxes also become time-dependent.

The paper is organized as follows. In section \ref{sec2} we review the KL model and propose a new model with $W_0=0$ and two vacua, one of them being parametrically tunable to be AdS or Minskowski. In section \ref{sec3}, we study the lifetime of an uplifted dS vacuum in the KL and in our model, and show that their lifetimes are not compatible with the TCC. In section \ref{sec4}, we discuss whether time-independent fluxes could be allowed or not once we take all possible quantum corrections into account. The analysis in section \ref{sec4} is presented from M-theory point of view, because most of the IIB results are easier to handle from an eleven-dimensional perspective. In M-theory, these quantum corrections include the perturbative and non-perturbative, as well as non-local and topological, pieces. Using these ingredients,
we explore the consistency of the model with $W_0=0$ and the uplift of an AdS to dS vacuum from the dual M-theory side. In section \ref{sec5} we conclude with the possibility that four-dimensional dS space may be realized in a controlled laboratory as a Glauber-Sudarshan state even in the presence of the entire tower of $\alpha'$ corrections and time-dependent dilaton and with a discussion about non-perturbative backgrounds and their relevance for duality invariant setups. 

\section{Purely non-perturbative AdS vacua and uplift to dS}
\label{sec2}

In this section, we will revisit and generalize the KL model in order to get two AdS minima with only non-perturbative terms in the superpotential.

The idea behind the KKLT scenario was to look for a supersymmetric AdS vacuum once non-perturbative corrections were taken into account, which was then uplifted to a dS minima by adding a positive energy density from $\overline{\rm D3}$ branes. However, this model was soon generalized to have more than one minimum which does not occur at large (negative) values of the effective potential. This is known as the Kallosh-Linde (KL) \cite{Kallosh:2004yh} model, in which one assumes the standard no-scale tree-level K\"ahler potential and a racetrack form of the superpotential. A feature of this model was to stabilize the volume modulus in the susy Minkowski vacuum.

More concretely, we consider the bosonic sector of $d=4$ and $\mathcal{N}=1$ sugra theory with superpotential and no-scale K\"ahler potential given by:
\begin{equation}\label{themodel}
    W(\rho) =W_0 + \sum_j^N A_j e^{i a_j \rho}\,, \quad K(\rho, \Bar{\rho}) = -3 \ln(-i(\rho-\Bar{\rho})) \,.
\end{equation}
It is a further generalization of the KL model mentioned above, which can be recovered from above once we set $N=2$, with $N$ non-perturbative terms for generality. In order to avoid the obstructions mentioned in \cite{Sethi:2017phn}, we shall set $W_0 =0$ in the above expression for the superpotential, such that the solution respects susy before introducing non-perturbative corrections, as has been previously done in \cite{Blanco-Pillado:2018xyn}. This point requires further clarification which shall be elaborated upon later on. Moreover, by the virtue of being of the racetrack type, this form of the superpotential is also free from anti-brane backreaction constraints, as argued in \cite{Danielsson:2018ztv}. In conclusion, for our model, we choose $j =1,2,3$ along with $W_0=0$, so that we get two minima. On choosing one of the susy minima to be a Minkowski one, further stabilizes the scenario \cite{BlancoPillado:2005fn}.

The idea is to then uplift the shallower of the two minima to a dS minimum thereby now opening a new channel of decay from the dS minima to the (remaining) AdS one, thereby making the metastable dS vacua short-lived (following \cite{Ceresole:2006iq}). When we finally turn to numerics, it seems that it is impossible to make the lifetime of this dS vacua to be within the bound as allowed by the TCC. 

\subsection{Generalizing the KL model}
For the model (\ref{themodel}), the covariant K\"ahler derivative is:
\begin{align}
    D_{\rho}W &\equiv  \partial_\rho W + W \partial_\rho K\nonumber\\
    &= -\frac{3 W_0}{\rho -\Bar{\rho}}+ \sum_j A_j \left(i a_j - \frac{3}{\rho -\Bar{\rho}}\right)e^{i a_j \rho}\,,
\end{align}
while the K\"ahler metric is given by:
\begin{equation}
    K_{\rho\Bar{\rho}} \equiv \partial_{\rho}\partial_{\bar{\rho}} K = -\frac{3}{(\rho- \Bar{\rho})^2}\,.
\end{equation}
The potential for the complex scalar field $\rho = \phi + i \sigma$ is given by (setting $M_{\text{Pl}}=1$):
\begin{align}
    V(\rho) &= e^K\left[K^{\rho\Bar{\rho}}D_{\rho}W D_{\Bar{\rho}}\overline{W}- 3 |W|^2\right]\\
    &= \frac{1}{i(\rho - \Bar{\rho})^3}\left[-\frac{(\rho - \Bar{\rho})^2}{3}\left| -\frac{3 W_0}{\rho -\Bar{\rho}} + \sum_j A_j \left(i a_j - \frac{3}{\rho - \bar{\rho}}\right)e^{i a_j \rho}\right|^2 -3\left|W_0 + \sum_j A_j e^{i a_j \rho}\right|^2\right]\,.\nonumber
\end{align}
Let us turn off the axion, so that  we have $\rho = i \sigma$. In this case, the potential for the volume $\sigma$ of the 4-cycle is (assuming $A_i$ and $a_i$ to be real):
\begin{equation}
    V(\sigma) = \frac{1}{6\sigma}\sum_{i,j}A_i A_j \left(a_i a_j + \frac{3}{\sigma}a_j\right)e^{-\sigma(a_i + a_j)} + \frac{W_0}{2\sigma^2}\sum_i A_i a_i e^{-a_i \sigma}.
\end{equation}
The condition for a supersymmetric vacuum (that comes from the susy transformations of the fermions) is:
\begin{equation}\label{susyvacuum}
    (D_\rho W)(\sigma_0) = 0 \implies W_0 + \sum_j A_j\left(1+ a_j\frac{2\sigma_0}{3} \right)e^{-a_j\sigma_0}=0.
\end{equation}
If, on top of that, we impose the minimum to be Minkowski, we require that:
\begin{equation}\label{susymink1}
    W(\sigma_0) = 0 \implies W_0 + \sum_j A_j e^{-a_j \sigma_0} = 0.
\end{equation}
Combining this last equation with (\ref{susyvacuum}), we get:
\begin{equation}\label{susymink2}
    \sum_j A_j a_j e^{-a_j\sigma_0}=0,
\end{equation}
assuming $\sigma_0 \neq 0$ for consistency. In fact, the last equation can be used to find the value of $\sigma_0$ explicitly, at least in the case of the KL model ($A_j = 0$ for $j>2$) or for models with $a_1 = a$ and $a_i = a_2$ ($i>1$), in which case instead of a sum with different exponential factors in each term we get only two terms with different exponentials:
\begin{equation}\label{sigmazeroforWnonzero}
    A_1a_1 e^{-a_1 \sigma_0} + \left(\sum_{j\neq 1} A_j\right)a_2e^{-a_2\sigma_0} = 0 \implies \sigma_0 = \frac{1}{a_1 - a_2}\ln \left(- \frac{A_1 a_1}{a_2\sum_{j\neq 1} A_j}\right).
\end{equation}
Now, suppose we want this supersymmetric Minkowski vacuum to be AdS. We need $V(\sigma_0)<0$ and so condition (\ref{susymink1}) to be violated. In other words, we are still going to impose (\ref{susyvacuum}) but, instead of (\ref{susymink1}), we shall demand:
\begin{equation}
    W_0 + \sum_j A_j e^{-a_j \sigma_0} = \delta = \delta W_0\,, 
\end{equation}
where we used the notation $\delta W_0$ because the non-vanishing of $W(\sigma_0)$ can be interpreted as a change in $W_0$. This procedure had already been mentioned in \cite{Kallosh:2019axr}, in order to get an AdS minimum through a small change in $W_0$. Note that we need to assume that $\delta W_0 \ll W_0$ in order for $\sigma_0$ to have approximately the same value before the change in the parameter. Thus, using (\ref{susyvacuum}), we get:
\begin{equation}
    \sum_j A_j a_j e^{-a_j\sigma_0}=-\frac{3}{2\sigma_0}\delta W_0\,.
\end{equation}
Using these conditions, the potential evaluated at $\sigma_0$ turns out to be:
\begin{equation}
    V(\sigma_0) = -\frac{1}{6\sigma_0}\left(\sum_j A_j a_j e^{-a_j\sigma_0}\right)^2 \approx -\frac{3}{8\sigma_0^3}\delta W_0^2<0\,,
\end{equation}
so that, as expected, we get an AdS vacuum. 

Now, let us consider the interesting case of $W_0 = 0$. Let us emphasize right at the beginning that this does not imply that we do not have fluxes stabilizing the complex structure moduli, but rather that such fluxes are tuned in such a way so that once we integrate out the complex structure, we get $W_0=0$. In other words, if we denote the general GVW superpotential as $W_{\text{GVW}}$ we are demanding that $W_0 \equiv \langle W_{\text{GKP}} \rangle$ = 0 and are looking for solutions with $W_0=0$ in the vacuum at the stabilized values of the complex structure moduli. At this point, a natural question to ask is if it is indeed possible to stabilize the complex structure moduli if $W_0=0$? And the answer to this question is yes, as has already been noted in \cite{Blanco-Pillado:2018xyn}. Following \cite{Curio:2006ea, Denef:2004dm}, keeping only the complex structure (before adding the non-perturbative terms), one can find appropriate flux numbers such that there exists a vacua with a vanishing tree level superpotential.  There have been similar models studied in the past demonstrating that one can stabilize the axion-dilaton and the complex structure moduli while maintaining $W_0=0$ \cite{Giryavets:2003vd}. In a similar spirit, our vacua are constructed purely from the non-perturbative gaugino-condenstation terms. As a result, we do not have susy breaking terms even before adding in the non-perturbative contributions, thereby side-stepping the obstructions mentioned in \cite{Sethi:2017phn}.

Having taken this detour to explain why we are interested in taking $W_0=0$, let us return to our calculation. In this case, the conditions for the supersymmetric  Minkowski vacuum reduces to:
\begin{align}\label{susyminknoW1}
    (D_\rho W)(\sigma_0) &= 0 \implies \sum_j A_j\left(1+ a_j\frac{2\sigma_0}{3} \right)e^{-a_j\sigma_0}=0\,,\\
    W(\sigma_0) &= 0 \implies \sum_j A_j e^{-a_j \sigma_0} = 0\,. \label{susyminknoW2}
\end{align}
At first sight, it seems that we could use equation (\ref{sigmazeroforWnonzero}) to find the value of $\sigma_0$. This is actually not true, as can be seen by plugging (\ref{sigmazeroforWnonzero}) into (\ref{susyminknoW2}), that would imply that we have $a_1= a_2$ and so all $a_i$ are equal. This already indicates that (\ref{sigmazeroforWnonzero}) cannot be used in this case. We need to combine equations (\ref{susyminknoW1}) and (\ref{susyminknoW2}) to get (\ref{susymink2}) again, which under the assumption that $a_j =a_2$ for $j>1$ implies that:
\begin{equation}\label{sigmazeroforWzero}
    e^{(a_1-a_2)\sigma_0} =-\frac{A_1 a_1}{a_2 \sum_{j\neq 1}A_j}\,.
\end{equation}
On inserting this result, in turn, into (\ref{susyminknoW2}), we find $a_1 = a_2$.  Finally, using (\ref{sigmazeroforWzero}) implies that:
\begin{equation}
    \sum_{j}A_j = 0\,.
\end{equation}
This, however, means that we get no constraint on $\sigma_0$! In fact, it can be easily shown that the full potential $V(\sigma)$ vanishes after imposing $a_i = a$ for all $i$ and $A_1 + \cdots+ A_N = 0$, since $W$ itself vanishes. Thus, under these conditions, \emph{we have a flat direction for the volume modulus.}

Thus, we need to violate our assumption that $a_j = a_2$ (for any $j>1$) in order to get a non-trivial potential. In fact, consider the general conditions we have for a Minkowski vacuum, equations (\ref{susymink2}) and (\ref{susyminknoW2}). With two subsets of equal $a_i$ these equations have the following schematic structure:
\begin{align}
    A e^x + Be^{\alpha x} &=0\,,\\
    A e^x + B \alpha e^{\alpha x} &=0\,,
\end{align}
from which it is straightforward to see that the only solution is for $A = -B$, $\alpha =1$ for any value of $x$. This explains why we got the vanishing potential above. So, we need at least three subsets of equal $a_i$ to get a non-trivial potential. This shows that we need $N \geq 3$.

A glance at the structure of the full superpotential $W$ shows that, effectively, if we choose to work with $n$ subsets of equal $a_i$, then \emph{we only need to study the $N=n$ model}. More concretely, consider $n$ subsets with $a_{i_1} = \alpha_1$ for $i_1 \in [1,N_1]$, $a_{i_2} = \alpha_2$ for $i_2 \in [N_1 + 1, N_2]$, $\dots$ and $a_{i_n} = \alpha_n$ for $i_{n} \in [N_{n-1} +1, N]$. The conditions for having a  Minkowski vacuum becomes:
\begin{align}
    \alpha_{1} e^{-\alpha_1\sigma_0} \sum_{i_1} A_{i_1} + \alpha_2 e^{-\alpha_2 \sigma_0} \sum_{i_2} A_{i_2} + \cdots + \alpha_n e^{-\alpha_n\sigma_0} \sum_{i_n} A_{i_n} &=0\,,\\
    e^{-\alpha_1\sigma_0} \sum_{i_1} A_{i_1} + e^{-\alpha_2 \sigma_0} \sum_{i_2} A_{i_2} + \cdots + e^{-\alpha_n\sigma_0} \sum_{i_n} A_{i_n} &=0\,,
\end{align}
and has the same structure as in the model with $N=n$, with the one-loop determinant coefficients $A_i$ being replaced by sums of the coefficients of each subset. Combining the equations above, we get:
\begin{equation}
    e^{-\alpha_2 \sigma_0}(\alpha_2 - \alpha_1)\sum_{i_2} A_{i_2} + \cdots + e^{-\alpha_n\sigma_0}(\alpha_n - \alpha_1)\sum_{i_n}A_{i_n} = 0\,,
\end{equation}
such that apart from the $n=3$ case, it is a transcendental equation for $\sigma_0$.

In order not to add any unnecessary tuning, we will consider the case with the minimal number of 3 subsets. Note that this is the only case in which we could write a closed analytical expression for $\sigma_0$. In light of the previous discussion, we shall take $N=3$ and we will consider $a_1 \neq a_2\neq a_3$. As compared with the KL model, in this case we are dropping out a degree of freedom ($W_0$) from the superpotential but we are allowing an extra non-perturbative term to play its role. We can easily solve the condition (\ref{susyminknoW2}) by setting:
\begin{equation}\label{A1fornoW}
    A_1 = -\left(A_2 e^{(a_1-a_2)\sigma_0} + A_3 e^{(a_3 - a_2)\sigma_0}\right)\,.
\end{equation}
Now we can use equation (\ref{susymink2}) to find $\sigma_0$, as follows:
\begin{equation}
    \sigma_0 = \frac{1}{(a_3 - a_2)}\ln\left[-\frac{(a_3 - a_1) A_3}{(a_2 - a_1)A_2}\right]\,.
\end{equation}
There is another minimum which is AdS, obtained by imposing $W(\sigma_0) \neq 0$.

Now we can proceed as in the KL case, to perturb the parameters a little  to get two AdS vacua. This time we are going to use a small change in $A_1$ to promote the Minkowskian minimum to be an AdS one. So, instead of (\ref{susyminknoW2}), we impose:
\begin{equation}
    \sum_j A_j e^{-a_j \sigma_0} = \delta A_1 e^{-a_1 \sigma_0}\,,
\end{equation}
and using this into (\ref{susyminknoW1}), we get:
\begin{equation}
    \sum_j A_j a_j e^{-a_j \sigma_0} = - \frac{3}{2\sigma_0}\delta A_1 e^{-a_1 \sigma_0}\,.
\end{equation}
Evaluating the potential under such conditions, we have:
\begin{equation}
    V(\sigma_0) \approx -\frac{3}{8\sigma_0^3} (\delta A_1) ^2 e^{-2a_1 \sigma_0}<0\,,
\end{equation}
which indeed gives an AdS minimum. Notice that now we require that $\delta A_1/A_1 \ll 1$ in order for $\sigma_0$ to remain practically unchanged.

In order to get a dS vacuum, we assume an uplifting term in the potential from $\overline{\text{D3}}$ in the same lines as in the original KL model. Since the amount of uplift is parametrically smaller as compared with the KKLT case, the $W_0=0$ model potentially enjoys all the advantages discussed in \cite{Kallosh:2019axr} regarding the $\overline{\text{D3}}$ backreaction. Thus, only the shallower minima is uplifted to dS and the second one remains practically unchanged. Moreover, as discussed in the next section, the lifetime of the uplifted dS vacuum depends only on the values of the minima \textit{prior to the uplift}.

\section{Lifetime of the dS vacua}
\label{sec3}
\subsection{Estimates from analytic expressions}
We are interested in the lifetime of our non-perturbative dS minima since the TCC claims that, irrespective of the details of how one constructs a dS minima in string theory, its lifetime must satisfy the bound:
\begin{equation}\label{TCCbound}
    \tau < \frac{1}{H}\ln\left(\frac{1}{H}\right)  \approx 2.4 \times 10^{62}.
\end{equation}
where we used the Friedmann equation $3H^2 =V_0$ and the value of the cosmological constant today $V_0 \sim 10^{-120}$ to get the last relation. 

Models of stringy dS, such as the KKLT or LVS paradigms, produce one dS minima at a small positive value of $V_0$, as required by the cosmological constant today. As long as this value of $V_0$ is much smaller than the potential barrier which separates it from a Minkowski minimum at infinite volume, denoting the spontaneous decompactification of internal dimensions, we can estimate the lifetimes of such dS vacua by considering their quantum mechanical tunneling via Coleman-de Luccia instantons \cite{Westphal:2007xd}. The value of the lifetimes of all such vacua is estimated to be of the order of $\tau_{\rm dS} \sim e^{1/V_0} \sim e^{10^{120}}$ in Planck units. 

For our model with multiple AdS minima, there are new decay channels available for the dS vacua to tunnel into. This is the primary reason why the lifetime in this case would be smaller since we do not have to depend on the Minkowski vacuum at infinite volume to be the only steady state to decay into. However, we cannot use the standard analyses for calculating the lifetime of KKLT-like constructions \cite{Westphal:2007xd} to estimate the lifetime for our model. Instead, we refer to  \cite{Ceresole:2006iq}, in which it was shown that the lifetime of the dS vacuum that was uplifted from the first AdS vacuum (as in the KL model) is given by:
\begin{equation}\label{lifetimetwominima}
    \tau_{\text{dS}} = e^{\frac{24 \pi^2}{|V_0|}\frac{(C-1)^4}{C^2(2C-1)^2}},
\end{equation}
where $C^2 \equiv |V_1|/|V_0|$ with $V_0$ being the value of the potential at the quasi-Minkowski AdS vacuum and $V_1$ is its value at the second AdS vacuum (before the uplift). If we assume that $C \sim \sqrt{2}$, then the condition (\ref{TCCbound}) is satisfied when:
\begin{equation}\label{V0constraint}
    |V_0| \gtrsim 8.3 \times 10^{-3}.
\end{equation}
The expression \eqref{lifetimetwominima} deserves some comments. Firstly, the thin-wall approximation was extensively used to derive it which is well-justified in our case since the difference in the depth of the two minima is much smaller that the barrier between them \cite{Westphal:2007xd}. Secondly, it was implicitly assumed that on uplifting the shallower of the two minima, the depth of the other AdS minimum remains the same. Finally, it was approximated  that the tension of the bubble wall does not change on uplifting, although in practice the tension always decreases slightly during the uplift. Indeed, it is the decrease in the wall tension that allows the gravitation suppression to the formula to the tunneling amplitude to slowly decrease and allow for transitions between the uplifted vacua \cite{Ceresole:2006iq}. However, as it shall turn out, the numerical value of the transition amplitude shall not depend crucially on this assumption for our choice of parameters.

\begin{figure}[t]
\centering
\includegraphics[width =0.7\textwidth]{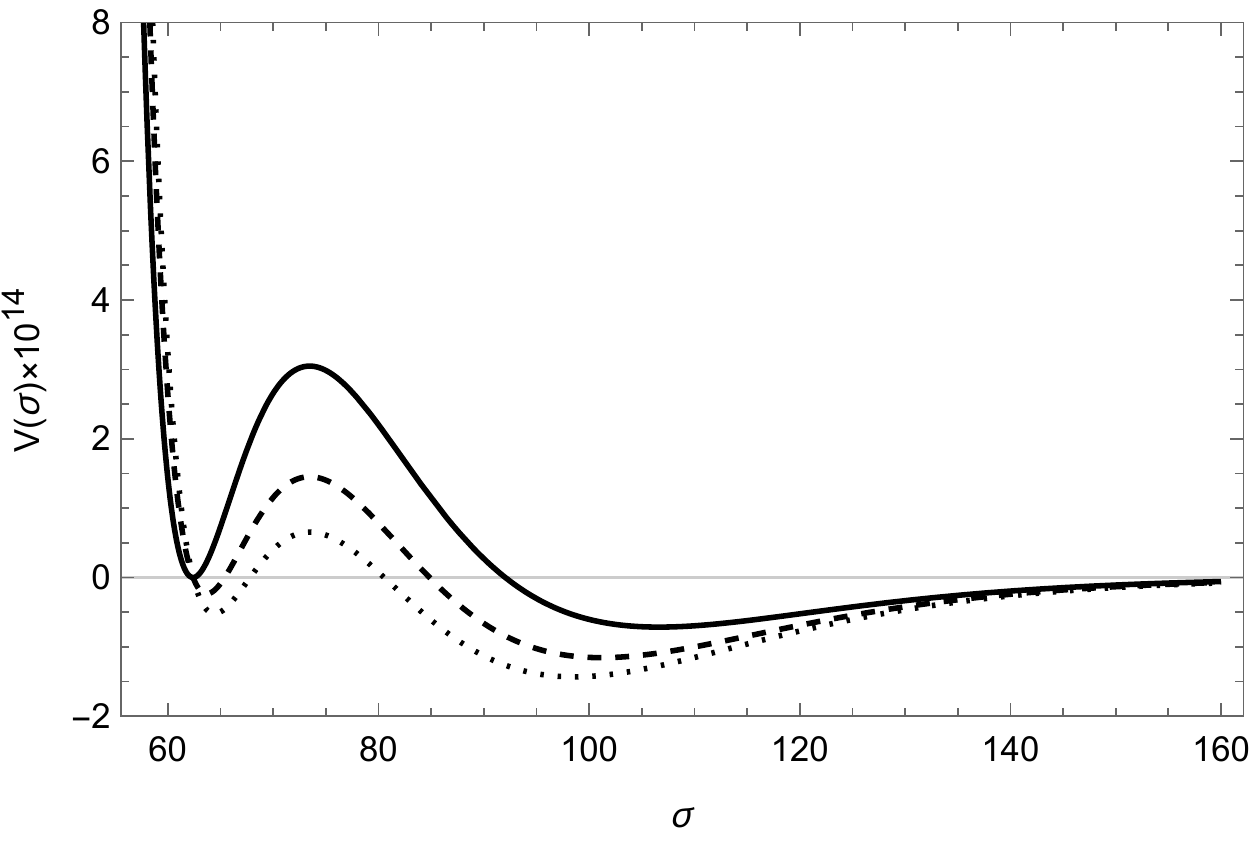}
\caption[]{The KL potentials for $A_1 = 1$, $A_2 = -1.03$, $a_1 = 2\pi/100$, $a_2 = 2\pi/99$ and $W_0 = -(1 + \delta)\sum_j A_j e^{-a_j\sigma_0}$ with $\sigma_0$ given by equation (\ref{sigmazeroforWnonzero}). The thick line corresponds to $\delta =0$, while $\delta = 0.2$ and $0.3$ corresponds to the dashed and dotted lines, respectively.}
\label{Figure1}
\end{figure}

To check whether we can tune the parameters of the models in the last section to satisfy this condition, consider for instance the the values of the parameters as in \cite{Kallosh:2004yh}, $A_1= 1$, $A_2=-1.03$, $W_0 = -2 \times 10^{-4}$ and $a_1$ and $a_2$ such that the two minima are at $\sigma_0 \approx 60$ and $\sigma_1 \approx 100$. In order to make estimates, let us assume though that $\sigma_0\approx 50 $, $\sigma_1 \approx 100$, $A_i\approx 1$ and $a_i \approx 10^{-1}$ such that $\sigma_i a_i >1$ (for consistency with neglecting high order instantons) and $|W_0| \sim 10^{-4}$. Then:
\begin{equation}
    V(\sigma_0) \approx -\frac{4}{3\sigma_0^3}\left(\frac{\delta W_0}{W_0}\right)^2 W_0^2 \approx -\left(\frac{\delta W_0}{W_0}\right)^2 \times 10^{-13},
\end{equation}
and so $\delta W_0/W_0$ should be greater than $10^{5}$, that is inconsistent.

For the generalized model with $W_0 =0$, we have:
\begin{equation}
    V(\sigma_0) \approx -\frac{3}{8\sigma^3}\left(\frac{\delta A_1}{A_1}\right)^2 A_1^2 e^{-2 a_1 \sigma_0} \approx - (\delta A_1)^2 \times 1.4 \times 10^{-10},
\end{equation}
and so the TCC bound is satisfied only if $\delta A_1$ is greater than $10^3$, which is also inconsistent with $\delta A_1 \ll A_1$.

Thus, it seems that for generic values of the parameters that are consistent with the sugra assumptions and first instanton contribution, the models are not consistent with the TCC

\subsection{Numerical results}

The estimates in the previous section depend on the value of $C$ that was fixed to be $\sqrt{2}$ in order to estimate (\ref{V0constraint}). In fact, it is the presence of the $C>1$ depedent factor in (\ref{lifetimetwominima}) that yields a life time orders of magnitude smaller than (\ref{KKLT_lifetime}). To have a small life time, we need to consider a large $|V_0|$ and/or a $C$ very close to 1. In this section we consider the precise $V_0$ and $V_1$ values for some fixed parameter and show that the generic estimates of the previous section does apply.

Firstly, we consider the KL model. A plot of the potential for this case is shown in {\bf Figure \ref{Figure1}}. Once the parameters are fixed such that we have an initially supersymmetric Minkoswkian vacuum, the value $V_1$ of the potential at the second (non-supersymmetric AdS) minimum is fixed. This shows that, generically, we cannot tune the value of $C$ to be what we want. The life-times for the dashed and dotted potentials in the figure turns out to be $\tau \approx e^{3.8\times 10^{15}}$ and $\tau \approx e^{5.81 \times 10^{14}}$, respectively. Note that a slight change in $W_0$ makes the life-time decrease by many orders in magnitude, but not enough to satisfy the bound (\ref{TCCbound}). Thus, as expected due to the results of \cite{Ceresole:2006iq} about their life time, the vacua in these examples are very long-lived and do not satisfy the TCC.

\begin{figure}[t]
\centering
\includegraphics[width =0.7\textwidth]{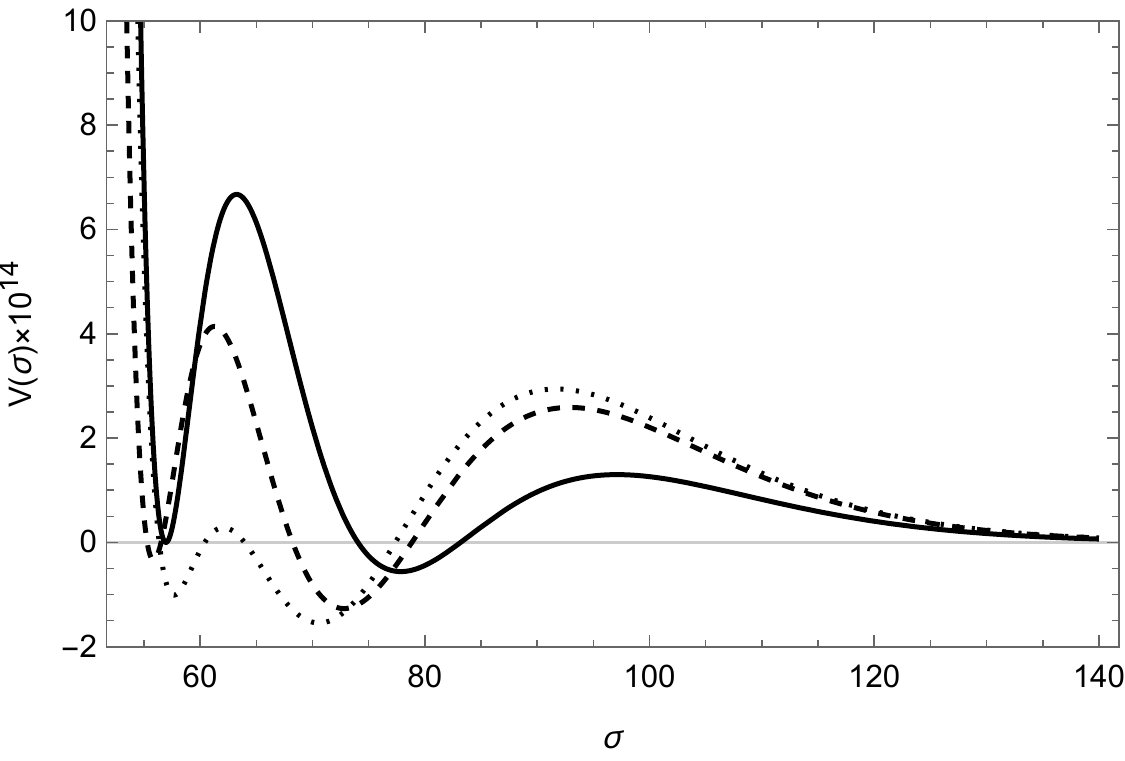}
\caption[]{The model with $W_0 =0$ and $N=3$ for $A_2 = 0.98$, $A_3 = -1.05$, $a_1 = 2\pi/100$, $a_2 = 2\pi/80$, $a_3 = 2\pi/70$ and $A_1 = -(1+ \delta)(A_2e^{(a_1 - a_2)\sigma_0} + A_3e^{(a_3 - a_2)\sigma_0})$ with $\sigma_0$ given by equation (\ref{sigmazeroforWzero}). The thick line corresponds to $\delta =0$, while $\delta = 0.007$ and $0.013$ corresponds to the dashed and dotted lines, respectively.}
\label{Figure2}
\end{figure}

For the model with $W_0 = 0$ and three non-perturbative terms in the superpontential, similar arguments apply. As can be seen in {\bf Figure \ref{Figure2}}, as we change the value of $A_1$ fixed by (\ref{A1fornoW}), there is no significant decrease in $C$, and $V_0$ cannot become very large without destroying the existence of two minima. A distinct feature of this case is the presence of an extra maximum as compared with the KL case. So the potentials approach zero from positive values as $\sigma \rightarrow \infty$. The life time for the dashed and dotted potentials are $\tau \approx e^{1.78\times 10^{15}}$ and $\tau \approx e^{2.23\times 10^{13}}$, respectively.

\section{Stringy embedding of four-dimensional de Sitter vacua}
\label{sec4}

In the previous sections, we worked on the 4-d framework of susy preserving type IIB flux compactification and discussed how one may use non-perturbative corrections in the superpotential to construct a model with two AdS minima, one of them being uplifted to a de Sitter minimum. In the rest of this paper, we would like to step back and investigate if our setup is consistent with the higher dimensional equations of motion. 
For simplicity, we will concentrate on the model with one AdS minimum, as the extension to two AdS minima is straightforward.

\noindent The analysis carried out in section \ref{sec2} is based on four crucial assumptions:

\vskip.1in

\noindent $\bullet$ The superpotential is constructed out of time-independent ${\bf H}_3$ and 
${\bf F}_3$ fluxes.

\vskip.1in

\noindent $\bullet$ The quantum corrections to the superpotential, for example the gaugino condensate and the D3-instanton terms, also remain time-independent.

\vskip.1in

\noindent $\bullet$ The $W_0 = 0$ condition can be maintained even if we switch on the quantum corrections to the superpotential.

\vskip.1in

\noindent $\bullet$ No other higher order corrections, other than the ones discussed here, are important or deemed necessary for the dynamics of the uplift.

\vskip.1in

\noindent In this section we will critically examine each of these assumptions, when inherited by de Sitter after uplift, to see whether it is at all possible for four-dimensional de Sitter space to exist as a {\it vacuum} in type IIB string theory. For the sake of computational efficiency, we will analyze all the dynamics from M-theory, wherein all the type IIB fluxes may be neatly packaged as four-form G-fluxes, whereas the IIB axio-dilaton become part of the eleven-dimensional metric. The M-theory analysis also lends us another advantage: the eleven-dimensional supergravity action may be augmented by the addition of an infinite collection of local and non-local, including their perturbative and non-perturbative as well as the topological, quantum corrections. 

\subsection{Uplifting an anti de Sitter space to a de Sitter space \label{liftup}}

Let us start by studying a background of the form  ${\rm AdS}_4 \times \mathbb{M}_6$ in the type IIB side. This is a supersymmetric background, but a realization of such a background is not straightforward as one would not only have to specify what type of internal manifold $\mathbb{M}_6$ $-$ meaning it's topology and geometry $-$ could be allowed, but also what types of fluxes and quantum effects (if any) are required to support a background of this type. In fact we can even generalize the story by considering a background of the form: 
\bg\label{karintag}
ds^2 = {1\over \sqrt{h(y, x_2)}}\left(-dt^2 + dx_1^2 + dx_2^2 + dx_3^2\right) + \sqrt{h_1(y)} g^{(0)}_{\rm MN} 
dy^{\rm M} dy^{\rm N}, \nd
where $x_2$ can be taken to be the {\it radial} direction and the warp-factor $h(y, x_2)$ depends on both $y^m$, the coordinates of the internal space $\mathbb{M}_6$, and $x_2$. In the limit when $h(y, x_2) = h(y)$, this background would be dual to the M-theory {\it solitonic} background studied in \cite{desitter2, desitter3} and more recently in \cite{coherbeta}. On the other hand, when:
\bg\label{lumierep}
h(y, x_2) =  x_2^4 h_1(y), \nd 
the background becomes ${\rm AdS}_4 \times \mathbb{M}_6$, where $\mathbb{M}_6$ is   circle fibration over a squashed Sasaki-Einstein manifold \cite{luest}. We want the circle fibration to degenerate locally over the Sasaki-Einstein base such that there is a two-cycle locally. In that case $\mathbb{M}_6$ decomposes as 
${\cal M}_4 \times {\cal M}_2$, converting \eqref{karintag} to the following:
\bg\label{preacher}
ds^2 = {1\over x_2^2\sqrt{h_1(y)}}\left(-dt^2 + dx_1^2 + dx_2^2 + dx_3^2\right) 
+ \sqrt{h_1(y)} \Big(g^{(0)}_{mn} dy^{m} dy^{n} + g^{(0)}_{\alpha\beta} dy^{\alpha} dy^{\beta}\Big), \nd
where $(m, n) \in {\cal M}_4$ and $(\alpha, \beta) \in {\cal M}_2$. The usefulness of such a decomposition of the internal geometry is already discussed in \cite{desitter2}, so we won't go into details here. The difference compared to \cite{desitter2} is that this decomposition is only local, but more importantly, the internal manifold can now have a ${\rm SU}(3) \times {\rm SU}(3)$ structure along-with {\it all} type IIB fluxes switched on \cite{luest}. 

Another key difference from \cite{desitter2, coherbeta} is that, due to the explicit $x_2$ dependence of the space-time metric, the M-theory uplift of this background will be slightly different. We have lost the full 
$\mathbb{R}^{3, 1}$ isometries, but we could still compactify the $x_3$ direction and uplift the resulting configuration to M-theory. This will be discussed soon, but the main aim of our analysis is to answer the following question: Under what condition(s) can the background \eqref{preacher} be converted to the following background:

{\footnotesize
\bg\label{iibmet}
{\mathrm d}s^2 =  {1 \over \Lambda {\rm H}^2(y) \vert t\vert^2}\left(-{\mathrm d}t^2 + {\mathrm d}x_1^2 + {\mathrm d}x_2^2 + {\mathrm d}x_3^2\right)
+ {\rm H}^2(y)\Big({\rm F}_1(t)  g_{\alpha\beta}(y) {\mathrm d}y^\alpha {\mathrm d}y^\beta + {\rm F}_2(t) 
g_{mn}(y) {\mathrm d}y^m {\mathrm d}y^n\Big), \nd}
where ${\rm H}(y)$ is now the warp-factor and $F_i(t)$ are the time-dependent coefficients constrained by the relation ${\rm F}_1(t) {\rm F}_2^2(t) = 1$ to keep the four-dimensional Newton's constant time-independent.

Comparing \eqref{preacher} with \eqref{iibmet} we see that the $x_2$ dependence of the space-time metric has disappeared from \eqref{iibmet}, and is replaced by $\Lambda\vert t\vert^2$. Clearly there is no simple 
coordinate transformation one could do to \eqref{preacher} to recover \eqref{iibmet}, so that path(s) leading from \eqref{preacher} to \eqref{iibmet} have to be more non-trivial\footnote{Transformations like 
$x_2 \to it, t \to i x_2$ and $\Lambda \to -\Lambda$ do not take us from \eqref{preacher} to \eqref{iibmet}, or vice versa.}. Additionally, \eqref{preacher} could either be supersymmetric or non-supersymmetric, both depending on the choices of fluxes and warp-factor, but \eqref{iibmet} has no supersymmetric limit. The inherent time dependences of the metric components do not allow for a closed supersymmetric algebra to be realized here. What about the fluxes supporting the background \eqref{iibmet}? In \cite{desitter2, desitter3, coherbeta} we have argued that the fluxes should have time-dependences to allow for a four-dimensional EFT description to be valid here. It is easy to show, from the M-theory uplift of \eqref{iibmet}, the localized fluxes of the form 
${\bf G}_{{\rm MN}ab}$, as well as the space-time flux components ${\bf G}_{0ij{\rm M}}$, where $({\rm M, N}) \in {\cal M}_4 \times {\cal M}_2$, $(a, b) \in {\mathbb{T}^2\over {\cal G}}$ and 
$(0, i, j) \in {\bf R}^{2, 1}$, have to remain time-dependent otherwise we will clash with the $g_s$  and the ${\rm M}_p$ hierarchies where $g_s$ is the corresponding type IIA string coupling.
The other G-flux components, for example ${\bf G}_{\rm MNPQ}$ and  ${\bf G}_{{\rm MNP}a}$, were also taken to be time-dependent in \cite{desitter2, desitter3, coherbeta}, and this fitted rather well with all the Einstein and the G-flux EOMs. The fluxes supporting the background \eqref{karintag}, on the other hand, 
are all time-{\it independent}\footnote{There are multiple way to show this and one of the simplest way would be to use the torsion classes. The latter automatically solves the supersymmetry conditions, in addition to giving the explicit background fluxes (see \cite{luest} for details)}.
A question that we want to ask here is whether the two set of G-flux components,  ${\bf G}_{\rm MNPQ}$ and  ${\bf G}_{{\rm MNP}a}$, could be kept time-{\it independent} without violating the flux EOMs or the hierarchies.  

At a face value, this might seem possible. At least the $g_s$ scalings of the perturbative quantum corrections, as detailed first in \cite{desitter2}, seem to allow this. This may be seen by first lifting the metric 
\eqref{iibmet} to M-theory as:

{\footnotesize
\bg\label{mup}  
{\mathrm d}s^2 = g_s^{-8/3} \eta_{\mu\nu} {\mathrm d}x^\mu {\mathrm d}x^\nu + g_s^{-2/3} {\rm H}^2(y)\Big({\rm F}_1(g_s) g_{\alpha\beta} {\mathrm d}y^\alpha {\mathrm d}y^\beta
+ {\rm F}_2(g_s) g_{mn} {\mathrm d}y^{m} {\mathrm d}y^{n}\Big) + g_s^{4/3} 
\vert {\mathrm d}z \vert^2, \nd}
where the type IIA coupling $g_s$ is given by $g_s = g_b \sqrt{\Lambda} \vert t \vert {\rm H}(y)$ with $g_b$ being the {\it constant} type IIB coupling\footnote{Recall from \cite{desitter2, desitter3, coherbeta},  we are at the constant coupling limit of F-theory \cite{DM}, so at least the dilaton is a constant.}
and $dz \equiv dx_3 + i dx_{11}$. The additional warp-factors ${\rm F}_i(g_s)$ are defined as functions of $g_s$ in the following way:
\bg\label{saboot}
 {\rm F}_i(g_s) = \sum_{k\ge 0} {\rm C}^{(i)}_k \left({g_s\over {\rm H}}\right)^{2k/3},~~~
 \sum_{\{k_i\} \ge 0} {\rm C}^{(1)}_{k_1} {\rm C}^{(2)}_{k_2} {\rm C}^{(2)}_{k_3}
 \left({g_s\over {\rm H}}\right)^{2(k_1 + k_2 + k_3)/3} = 1, \nd 
with constant  ${\rm C}^{(i)}_k$ and the latter condition appearing from ${\rm F}_1(g_s) {\rm F}^2_2(g_s) = 1$.
Such a metric configuration, with the G-flux choices listed earlier, can only be supported in the presence of quantum terms. The quantum terms, which we will simply write as  
$\mathbb{Q}_{\rm T}^{(\{l_i\}, n_i)}$ to comply with the notations used in \cite{desitter2, desitter3, coherbeta},
can be of the form local and non-local type, which may be further classified into perturbative, non-perturbative and topological quantum terms. The perturbative and the topological quantum terms have been elaborated in \cite{desitter2}, while the non-perturbative ones are discussed in \cite{coherbeta}. They all appear, at least in the Einstein's equations, as contributions to the energy-momentum tensors in the following way:
\bg\label{enermom}
\mathbb{T}_{\rm MN} \equiv  a_1 {\bf g}_{\rm MN} \mathbb{Q}_{\rm T}^{(\{l_i\}, n_i)} +  
{\partial \mathbb{Q}_{\rm T}^{(\{l_i\}, n_i)}\over \partial {\bf g}^{\rm MN}}, \nd
where $a_1$ is a constant that only depends on the space-time dimensions, and $({\rm M, N}) \in {\cal M}_4 \times {\cal M}_2 \times {\mathbb{T}^2\over {\cal G}}$, instead of just being restricted to the base  
${\cal M}_4 \times {\cal M}_2$ in the type IIB side. Using these ingredients we are ready to answer, in the following sections, whether certain G-flux components may remain time-independent.                                                       

\begin{table}[tb]  
 \begin{center}
\renewcommand{\arraystretch}{1.5}
\begin{tabular}{|c||c||c||c|}\hline Riemann tensors  & Components & 
$g_s$ expansions \\ \hline\hline
${\bf R}_{\rm MNPQ}$ &  ${\bf R}_{mnpq}, {\bf R}_{mnp\alpha}, {\bf R}_{mn\alpha\beta}, {\bf R}_{m\alpha\alpha\beta}, {\bf R}_{\alpha\beta\alpha\beta}$ &  ${}^{\Sigma}_{{}_{k \ge 0}} {\rm R}^{(k)}_{\rm MNPQ} 
\left({g_s\over {\rm H}}\right)^{2(k-1)/3}$ \\ \hline 
${\bf R}_{{\rm MN}ab}$ &  ${\bf R}_{mnab}, {\bf R}_{m\alpha ab}, 
{\bf R}_{\alpha\beta ab}$ &  ${}^{\Sigma}_{{}_{k \ge 0}} {\rm R}^{(k)}_{{\rm MN}ab} 
\left({g_s\over {\rm H}}\right)^{2(k + 2)/3}$ \\ \hline 
${\bf R}_{abab}$ &  ${\bf R}_{abab}$ & ${}^{\Sigma}_{{}_{k \ge 0}} {\rm R}^{(k)}_{abab} 
\left({g_s\over {\rm H}}\right)^{2(k + 5)/3}$ \\ \hline 
${\bf R}_{{\rm MNP}0}$ &  ${\bf R}_{mnp0}, {\bf R}_{mn\alpha 0}, {\bf R}_{m\alpha\beta 0}, 
{\bf R}_{0\alpha\alpha\beta}$ & ${}^{\Sigma}_{{}_{k \ge 0}} {\rm R}^{(k)}_{{\rm MNP}0} 
\left({g_s\over {\rm H}}\right)^{(2k - 5)/3}$ \\ \hline 
${\bf R}_{{\rm MN}\mu\nu}$ &  ${\bf R}_{mnij}, {\bf R}_{m\alpha ij}, {\bf R}_{\alpha\beta ij}, {\bf R}_{0m0n}, {\bf R}_{0\alpha 0\beta}, {\bf R}_{0m 0\alpha}$
 &  ${}^{\Sigma}_{{}_{k \ge 0}} {\rm R}^{(k)}_{{\rm MN}\mu\nu} 
\left({g_s\over {\rm H}}\right)^{2(k - 4)/3}$ \\ \hline 
${\bf R}_{{\rm M}0ij}$ &  ${\bf R}_{m0ij}, {\bf R}_{\alpha 0ij}$
 & ${}^{\Sigma}_{{}_{k \ge 0}} {\rm R}^{(k)}_{{\rm M}0ij} 
\left({g_s\over {\rm H}}\right)^{(2k - 11)/3}$ \\ \hline  
${\bf R}_{\mu\nu\mu\nu}$ &  ${\bf R}_{ijij}, {\bf R}_{0i0j}$
 & ${}^{\Sigma}_{{}_{k \ge 0}} {\rm R}^{(k)}_{\mu\nu\mu\nu} 
\left({g_s\over {\rm H}}\right)^{2(k - 7)/3}$ \\ \hline   
${\bf R}_{0{\rm M} ab}$ &  ${\bf R}_{0mab}, {\bf R}_{0\alpha ab}$
 &  ${}^{\Sigma}_{{}_{k \ge 0}} {\rm R}^{(k)}_{0{\rm M}ab} 
\left({g_s\over {\rm H}}\right)^{(2k +1)/3}$ \\ \hline   
${\bf R}_{\mu\nu ab}$ &  ${\bf R}_{abij}, {\bf R}_{0a 0 b}$
 &  ${}^{\Sigma}_{{}_{k \ge 0}} {\rm R}^{(k)}_{\mu\nu ab} 
\left({g_s\over {\rm H}}\right)^{2(k - 1)/3}$ \\ \hline    
 \end{tabular}
\renewcommand{\arraystretch}{1}
\end{center}
 \caption[]{The ${g_s}$ expansions of the various curvature tensors associated with the metric \eqref{mup}. 
 Here $({\rm M, N}) \in {\cal M}_4 \times {\cal M}_2$ with a finer sub-division of $(m, n) \in {\cal M}_4$ and 
 $(\alpha, \beta) \in {\cal M}_2$. The other components are $(a, b) \in {\mathbb{T}^2\over {\cal G}}$ and 
 $(\mu, \nu) \in \mathbb{R}^{2, 1}$. The curvature tensor ${\rm R}^{(k)}_{\rm N_1N_2 N_3 N_4}  = 
 {\rm R}^{(k)}_{\rm N_1N_2 N_3 N_4}(y)$ where $y^m \in {\cal M}_4 \times {\cal M}_2$ and 
 $k \in {\mathbb{Z}\over 2}$.
These curvature tensors form the essential ingredients of the quantum terms \eqref{phingsha3}.} 
  \label{firzas}
 \end{table}

\subsection{Quantum series, curvature tensors and G-flux components \label{qotom}}

Existence of a de Sitter space relies heavily on the existence of the quantum terms. The quantum terms can be perturbative or non-perturbative, including non-local and topological, and in the following we will start with the perturbative terms constructed out of curvatures and flux components.
The various curvature and the flux components are collected in {\bf Table \ref{firzas}} and {\bf Table \ref{firzas2}} along-with their dominant $g_s$ scalings. We can put these together to determine the functional form for $\mathbb{Q}_{\rm T}^{(\{l_i\}, n_i)}$ as discussed in \cite{desitter2, desitter3} and \cite{coherbeta}. For us we can express it as:
\bg\label{phingsha3}
\mathbb{Q}_{\rm T}^{(\{l_i\}, n_i)} &= & \left[{\bf g}^{-1}\right] \prod_{i = 0}^3 \left[\partial\right]^{n_i} 
\prod_{{\rm k} = 1}^{27} \left({\bf R}_{\rm A_k B_k C_k D_k}\right)^{l_{\rm k}} \prod_{{\rm r} = 28}^{38} 
\left({\bf G}_{\rm A_r B_r C_r D_r}\right)^{l_{\rm r}}\nonumber\\
& = & {\bf g}^{m_i m'_i}.... {\bf g}^{j_k j'_k} 
\{\partial_m^{n_1}\} \{\partial_\alpha^{n_2}\} \{\partial_a^{n_3}\}\{\partial_0^{n_0}\}
\left({\bf R}_{mnpq}\right)^{l_1} \left({\bf R}_{abab}\right)^{l_2}\left({\bf R}_{pqab}\right)^{l_3}\left({\bf R}_{\alpha a b \beta}\right)^{l_4} \nonumber\\
&\times& \left({\bf R}_{\alpha\beta mn}\right)^{l_5}\left({\bf R}_{\alpha\beta\alpha\beta}\right)^{l_6}
\left({\bf R}_{ijij}\right)^{l_7}\left({\bf R}_{ijmn}\right)^{l_8}\left({\bf R}_{iajb}\right)^{l_9}
\left({\bf R}_{i\alpha j \beta}\right)^{l_{10}}\left({\bf R}_{0mnp}\right)^{l_{11}}
\nonumber\\
& \times & \left({\bf R}_{0m0n}\right)^{l_{12}}\left({\bf R}_{0i0j}\right)^{l_{13}}\left({\bf R}_{0a0b}\right)^{l_{14}}\left({\bf R}_{0\alpha 0\beta}\right)^{l_{15}}
\left({\bf R}_{0\alpha\beta m}\right)^{l_{16}}\left({\bf R}_{0abm}\right)^{l_{17}}\left({\bf R}_{0ijm}\right)^{l_{18}}
\nonumber\\
& \times & \left({\bf R}_{mnp\alpha}\right)^{l_{19}}\left({\bf R}_{m\alpha ab}\right)^{l_{20}}
\left({\bf R}_{m\alpha\alpha\beta}\right)^{l_{21}}\left({\bf R}_{m\alpha ij}\right)^{l_{22}}
\left({\bf R}_{0mn \alpha}\right)^{l_{23}}\left({\bf R}_{0m0\alpha}\right)^{l_{24}}
\left({\bf R}_{0\alpha\beta\alpha}\right)^{l_{25}}
\nonumber\\
&\times& \left({\bf R}_{0ab \alpha}\right)^{l_{26}}\left({\bf R}_{0ij\alpha}\right)^{l_{27}}
\left({\bf G}_{mnpq}\right)^{l_{28}}\left({\bf G}_{mnp\alpha}\right)^{l_{29}}
\left({\bf G}_{mnpa}\right)^{l_{30}}\left({\bf G}_{mn\alpha\beta}\right)^{l_{31}}
\left({\bf G}_{mn\alpha a}\right)^{l_{32}}\nonumber\\
&\times&\left({\bf G}_{m\alpha\beta a}\right)^{l_{33}}\left({\bf G}_{0ijm}\right)^{l_{34}} 
\left({\bf G}_{0ij\alpha}\right)^{l_{35}}
\left({\bf G}_{mnab}\right)^{l_{36}}\left({\bf G}_{ab\alpha\beta}\right)^{l_{37}}
\left({\bf G}_{m\alpha ab}\right)^{l_{38}}, \nd
where the 27 components of the Riemann tensors are listed in {\bf Table \ref{firzas}} and the 11 G-flux components are listed in {\bf Table \ref{firzas2}}. Each of these components are raised to the powers of 
$l_{\rm k}$ and $l_{\rm r}$ respectively along-with $n_i$ derivative actions, with $(0, 1, 2, 3)$ referring to the number of derivatives along time, ${\cal M}_4, {\cal M}_2$ and $x_{11}$ directions respectively. Finally everything is contracted by inverse metric components denoted here by $\left[{\bf g}^{-1}\right]$ symbolically.
This way $\mathbb{Q}_{\rm T}^{(\{l_i\}, n_i)}$ becomes a well-defined Lorentz-invariant quantity. The {\it full} perturbative quantum terms then are sum over all $(n_i, l_{\rm k}, l_{\rm r})$, and using this we can express our M-theory {\it effective} action in the following way:

{\footnotesize
\bg\label{sheela2}
\hskip-.1in{\bf S} &=& {\rm M}_p^9 \int d^{11} x \sqrt{-{\bf g}_{11}}\Big({\bf R}_{11} + {\bf G}_4 \wedge \ast {\bf G}_4 + 
{\bf C}_3 \wedge {\bf G}_4 \wedge {\bf G}_4 + {\rm M}_p^2 ~{\bf C}_3 \wedge \mathbb{Y}_8\Big)
-{\rm M}_p^{11}\sum_{k\ge 1} c_k\int d^{11}x\sqrt{-{\bf g}_{11}}\nonumber\\
&+& \sum_{\{l_i\}, n_i}  \int d^{11} x \sqrt{-{\bf g}_{11}}\left(
\mathbb{Q}_{\rm T}^{(\{l_i\}, n_0, n_1, n_2, n_3)}(y, g_s) \over {\rm M}_p^{\sigma(\{l_i\}, n_i) - 11}\right)
+ {\rm M}_p^3 \sum_{r = 1}^\infty \int d^3x \sqrt{-{\bf g}_3} ~ c_{(r)} \mathbb{W}^{(r)}(y, g_s)\nonumber\\
&-& {n_b {\rm T}_2\over 2}\int d^3\sigma \left\{\sqrt{-\gamma_{(2)}}\Big(\gamma^{\mu\nu}_{(2)} 
\partial_\mu X^{\rm M}\partial_\nu X^{\rm N} {\bf g}_{\rm MN} - 1\Big) + {1\over 3} \epsilon^{\mu\nu\rho} 
\partial_\mu X^{\rm M}\partial_\nu X^{\rm N}\partial_\rho X^P {\bf C}_{\rm MNP}\right\}\\
&+& \sum_{\{l_i\}, n_i} {\rm T}_7 \int d^7 \sigma \sqrt{-{\bf g}_7}\left[{\bf g}^{ab} ~\partial_a \partial_b 
\left({\hat{\mathbb{Q}}_{\rm T}^{(\{l_i\}, n_i)} (y, y^a, g_s) \over {\rm M}_p^{\sigma(\{l_i\}, n_i) - 7}}\right)
+ ~\Big(\bar{\bf\Psi}{\bf\Psi}\Big)^q  
\left({\widetilde{\mathbb{Q}}_{\rm T}^{(\{l_i\}, n_i)} (y, g_s) \over {\rm M}_p^{\sigma(\{l_i\}, n_i) - 7}}\right)\right],
\nonumber\\
& + & \sum_{\{l_i\}, n_i, k} {\rm M}_p^{11}\int d^{11} x \sqrt{-{\bf g}_{11}} \sum_{r = 1}^\infty  c_k~
{\rm exp}\Big[- k {\rm M}_p^6 \int d^6 y \sqrt{{\bf g}_6(y, g_s)}
~\mathbb{F}^{(r)}(x - y) \mathbb{W}^{(r - 1)}(y, g_s)\mathbb{V}_2(y, g_s)\Big], \nonumber \nd}
where the first line has the standard M-theory supergravity Lagrangian with $\mathbb{Y}_8$ forming the quartic order polynomial constructed out of curvature and G-flux two-forms (to be defined soon). The last entry, that appears as a volume factor, is there to cancel certain divergences which will be described below. The second line has the infinite number of perturbative interactions as defined in \eqref{phingsha3}, and the second term gives the non-local interactions that we write as:

{\footnotesize
\bg\label{romiran2} 
\mathbb{W}^{(r)}_{(\{l_i\}, n_i)}(y, g_s) &=& {\rm M}_p^8 \int d^8y' \sqrt{{\bf g}_8(y', g_s)} ~\mathbb{F}^{(r)}(y - y') \mathbb{W}^{(r-1)}_{(\{l_i\}, n_i)}(y', g_s)\\
& = & \int d^8 y' \sqrt{{\bf g}_8(y', g_s)}~\mathbb{F}^{(r)}(y - y').....\int d^8 y'' \sqrt{{\bf g}_8(y'', g_s)} \left({ \mathbb{F}^{(1)}(y - y'') 
\mathbb{Q}_{\rm T}^{(\{l_i\}, n_i)} (y'', g_s) \over {\rm M}_p^{\sigma(\{l_i\}, n_i) - 8r}}\right), \nonumber \nd}
which, as we see, is again constructed out of the perturbative terms in \eqref{phingsha3} but are connected by the non-locality functions  $\mathbb{F}^{(r)}(y - y')$, with $\sigma(\{l_i\}, n_i) \equiv \sigma_i$ being the ${\rm M}_p$ dimension. These functions become delta functions at low energies and at the supergravity level, so the non-local effects become invisible (see \cite{desitter2, desitter3} for more details on this). The third line is the action for $n_b$ number of well-separated M2-branes, although we should look for multiple M2-branes on top of each other (or being randomly distributed). There should also be interaction terms on the M2-branes, but we don't specify them here. The fourth line denotes the interactions on the uplifted IIA six-brane. This will dualize to the interaction terms on the IIB seven-brane. Again one should go with the non-abelian picture, but we avoid this to not over-complicate the system.  The interactions on the seven-brane are quantified by two terms: ${\hat{\mathbb{Q}}}_{\rm T}^{(\{l_i\}, n_i)} (y, y^a, g_s)$ and ${\widetilde{\mathbb{Q}}}_{\rm T}^{(\{l_i\}, n_i)} (y, g_s)$. The former is similar to \eqref{phingsha3} except for the explicit dependence of the flux components ${\bf G}_{{\rm MN}ab}$ on $y^b \equiv x_{11}$ and the seven-brane embedding\footnote{In fact the $y^b$ dependence appears explicitly from the normalizable form $\Omega_{ab}$ that is used to quantify the localized G-flux components ${\bf G}_{{\rm MN}ab}$. There is however no strong reason for it to depend on $y^b$ and in fact in \cite{coherbeta} we saw how the $y^\alpha$ and $g_s$ dependence of $\Omega_{ab}$ were more appropriate there. We could also simply make 
$\Omega_{ab} = \epsilon_{ab}$ and keep all $y^b$ dependence on the embedding itself.  These details may be gathered from \cite{coherbeta}, so we won't elaborate them here.}. The second term, 
${\widetilde{\mathbb{Q}}}_{\rm T}^{(\{l_i\}, n_i)} (y, g_s)$ incorporates the fermionic contributions on the dual seven-brane, and is similar to \eqref{phingsha3} with two key changes: one, some of the G-flux components are replaced as:
\bg\label{emmarob}
{\bf G}_{{\rm MN}ab} \to {\bf G}_{{\rm MN}ab} + {\rm M}_p  \bar{{\bf\Psi}} 
\Big(e_{11} {\Omega}_{{\rm MN}ab} + e_{12} \hat{\Omega}_{{\rm MN}ab}\Big){\bf\Psi} + ..., \nd
where $\bf{\Psi}$ is the world-volume fermionic degrees of freedom; ${\Omega}_{{\rm MN}ab}$ and 
$\hat{{\Omega}}_{{\rm MN}ab}$ are constructed out of the eleven-dimensional Gamma matrices etc.; and 
$e_{ij}$ are constants \cite{coherbeta}. Two, not all terms from \eqref{phingsha3} appear in the definition of 
${\widetilde{\mathbb{Q}}}_{\rm T}^{(\{l_i\}, n_i)} (y, g_s)$: terms from $l_{28}$ till $l_{35}$ are put to zero, after which \eqref{emmarob} is imposed on \eqref{phingsha3}. Finally, the last line involve an infinite sum over the exponentiated quantum terms, some of which would serve as BBS \cite{BBS} and KKLT \cite{KKLT} -type instanton like contributions, with $\mathbb{V}_2$ being the dimensionless warped volume of a two-manifold (see details in \cite{coherbeta}). This infinite series involve some divergences that is cancelled by the volume factor in the first line alluded to earlier.

\begin{table}[tb]  
 \begin{center}
\renewcommand{\arraystretch}{1.5}
\begin{tabular}{|c||c||c||c|}\hline G-fluxes  & Components & Dominant $g_s$ &
$g_s$ expansions \\ \hline\hline
${\bf G}_{\rm MNPQ}$ &  ${\bf G}_{mnpq}, {\bf G}_{mnp\alpha}, {\bf G}_{mn\alpha\beta}$ &  $\ge {1\over 3}$ & ${}^{\Sigma}_{{}_{k \ge 1/2}} {\cal G}^{(k)}_{\rm MNPQ} 
\left({g_s\over {\rm H}}\right)^{2k/3}$ \\ \hline 
${\bf G}_{{\rm MN}ab}$ &  ${\bf G}_{mnab}, {\bf G}_{m\alpha ab}, 
{\bf G}_{\alpha\beta ab}$ & 1 & ${}^{\Sigma}_{{}_{k \ge 3/2}} {\cal G}^{(k)}_{{\rm MN}ab} 
\left({g_s\over {\rm H}}\right)^{2k/3}$ \\ \hline 
${\bf G}_{{\rm MNP} a}$ &  ${\bf G}_{mnp a}, {\bf G}_{mn\alpha a}, {\bf G}_{m\alpha\beta a}$ & $\ge {1\over 3}$ &${}^{\Sigma}_{{}_{k \ge 1/2}} {\cal G}^{(k)}_{{\rm MNP}a} 
\left({g_s\over {\rm H}}\right)^{2k/3}$ \\ \hline 
${\bf G}_{0ij{\rm M}}$ &  ${\bf G}_{0ijm}, {\bf G}_{0ij\alpha}$ & $-4$ & ${}^{\Sigma}_{{}_{k \ge 0}} {\cal G}^{(k)}_{{\rm MNP}0} 
\left({g_s\over {\rm H}}\right)^{2(k - 6)/3}$ \\ \hline 
 \end{tabular}
\renewcommand{\arraystretch}{1}
\end{center}
 \caption[]{The ${g_s}$ scalings and expansions of the various G-flux components supporting the metric configuration \eqref{mup}. These G-flux components will also enter the perturbative quantum terms in 
 \eqref{phingsha3}. All the G-flux components ${\cal G}^{(k)}_{\rm N_1 N_2 N_3 N_4}(y)$ 
  depend on the coordinates of
 ${\cal M}_4 \times {\cal M}_2$ and $k \in {\mathbb{Z}\over 2}$.
 The numbers in the third column provide the dominant ${g_s\over H}$ scalings of the flux components.} 
  \label{firzas2}
 \end{table}

These are not the only terms comprising the action. There are the whole slew of eleven-dimensional fermionic interactions, amongst themselves, and with the bosonic degrees of freedom, that we 
didn't specify in 
\eqref{sheela2}. In some sense these terms would be a generalized version of 
${\widetilde{\mathbb{Q}}}_{\rm T}^{(\{l_i\}, n_i)} (y, g_s)$ in which all the curvature and the G-flux components from {\bf Tables \ref{firzas}} and
{\bf \ref{firzas2}} respectively are extended by the corresponding fermionic degrees of freedom as in 
\eqref{emmarob} (including the curvature tensors too), but now including intermediate fermions and inverse metric tensors, much like the ones discussed in section 4 of \cite{fermions}. This way, the  
generalized version of 
${\widetilde{\mathbb{Q}}}_{\rm T}^{(\{l_i\}, n_i)} (y, g_s)$ could, in principle, incorporate the infinite series of fermionic interactions that we want. 

Note that, in expressing the contents of {\bf Tables \ref{firzas}} and {\bf \ref{firzas2}}, we have not considered {\it all} possible choices of the curvature tensors and G-flux components. The reason is simple: all those curvature and G-flux components that do not appear in the two tables, {\it vanish} for the background 
\eqref{mup} as long as these components remain functions of the internal space $(y^m, y^\alpha) \in
{\cal M}_4 \times {\cal M}_2$ but are independent of the toroidal coordinates ${\mathbb{T}^2\over {\cal G}}$. The latter is crucial: $z \equiv x_3 + ix_{11}$ dependences will ruin the duality between M-theory and IIB, and that is also 
the reason for choosing $\Omega_{ab}$ $-$ the two-form used to quantify the localized G-flux components 
${\bf G}_{{\rm MN}ab}$ $-$ as functions of $y^\alpha$ in \cite{desitter3, coherbeta}. The relevant curvature and the G-flux components take the following form:
\bg\label{kirdunst}
&&{\bf R}_{\rm N_1N_2N_3N_4}(y, g_s) = \sum_{k \ge 0}{\rm R}^{(k)}_{\rm N_1 N_2 N_3 N_4}(y) 
\left({g_s\over {\rm H}}\right)^{l_{\{N_i\}} + 2k/3} \nonumber\\
&& {\bf G}_{\rm N_1N_2N_3N_4}(y, g_s) = \sum_{k}{\cal G}^{(k)}_{\rm N_1 N_2 N_3 N_4}(y) 
\left({g_s\over {\rm H}}\right)^{\tilde{l}_{\{N_i\}} + 2k/3}, \nd
where $l_{\{N_i\}}$ and $\tilde{l}_{\{N_i\}}$ may be read from {\bf Tables \ref{firzas}} and {\bf \ref{firzas2}}
respectively. The components not appearing in the two tables should be interpreted as being integrated out, although they would appear when we study fluctuations over the background \eqref{mup}. Plugging 
\eqref{kirdunst} in \eqref{phingsha3} tell us that it scales as $g_s^{\theta_{lkk'\hat{k}}}$, where:
\bg\label{ddeluca}
\hskip-.1in\theta_{lkk'\hat{k}} &\equiv& {2\over 3} \sum_{i = 1}^{27} (l_i + k_i) +{1\over 3}\left(\sum_{i = 0}^2 n_i - 2n_3 + l_{34} + 
l_{35}\right) + {2\over 3}\left(k'_1 + 2\right)\left(l_{28} + l_{29} + l_{31}\right) \nonumber\\
& + & {1\over 3}\left(2k'_2 + 1\right)\left(l_{30} + l_{32} + l_{33}\right) + {2\over 3}\left(k'_3 - 1\right)\left(l_{36} +
l_{37} + l_{38}\right) + \sum_{r = 1}^{38}\vert \hat{k}_r\vert l_{r},  \nd
where $l_i \in \mathbb{Z}$, $(k_i, k'_i, \hat{k}_i)  \in {\mathbb{Z}\over 2}$ and $n_i$ are the number of derivatives along temporal, ${\cal M}_4$, ${\cal M}_2$ and $y^b = x_{11}$ directions respectively. The 
$\vert \hat{k}_r\vert$ are related to 
${\rm F}_i(g_s)$, or products of them, and they appear from \eqref{saboot} but do not contribute at lowest order because $\vert \hat{k}_r\vert \ge 0$. 
It is clear from the above 
expression
that if we want to maintain a $g_s$ hierarchy, $k'_3 \ge {3\over 2}$ and $n_3 = 0$, although it appears that $k'_1 \ge 0$ and 
$k'_2 \ge 0$ may be possible.    
Since for these components $\tilde{l}_{\{N_i\}} = 0$, the {\it internal} G-flux components, 
other than ${\bf G}_{{\rm MN}ab}$, seem to allow for time-independent pieces ({\it i.e.} the $k = 0$ parts  of the G-flux components in \eqref{kirdunst}).  Could they be realized in our set-up?

The answer turns out to be intimately tied-up with the $\mathbb{Y}_8$ polynomial appearing in the first line of \eqref{sheela2}. For our case we can define the polynomial using the curvature and the G-flux two-forms, 
$\mathbb{R} \equiv {\bf R}^{a_ob_o}_{\rm N_1 N_2} {\bf M}_{a_ob_o} dy^{\rm N_1} \wedge dy^{\rm N_2}$ and $\mathbb{G} \equiv {\bf G}^{a_ob_o}_{\rm N_1 N_2} {\bf M}_{a_ob_o} dy^{\rm N_1} \wedge dy^{\rm N_2}$ respectively, in the following way:
\bg\label{kajalkir}
\mathbb{Y}_8 \equiv {\rm tr}  \left(\mathbb{R} + c_1\mathbb{G}\right)^4 
- {1\over 4} \left({\rm tr}\left(\mathbb{R} + c_2\mathbb{G}\right)^2\right)^2 + {\cal O}\left(\mathbb{G}^2\right)
= {\bf X}_8 + {\cal O}\left(\mathbb{R}^2 \mathbb{G}^2\right), \nd
where $(c_1, c_2)$ are constants\footnote{Independent of $(y, g_s)$ but dependent on ${\rm M}_p$.}, not necessarily equal, and in the limit they vanish, we recover the familiar 
${\bf X}_8$ polynomial. Such a form appears in both the flux EOMs as well as the anomaly cancellation conditions \cite{desitter2}, and therefore if it is independent of $g_s$ we can allow time-independent G-flux components ${\bf G}_{{\rm MNP}a}$ and ${\bf G}_{\rm MNPQ}$. In the following section we will evaluate the 
${\bf X}_8$ polynomial to see under what conditions it may be $g_s$ independent.


\subsection{Analysis of the curvatures and the ${\bf X}_8$ polynomial \label{X8}}

The analysis of the ${\bf X}_8$ polynomial requires us to first evaluate explicitly all the curvature tensors for the background \eqref{mup} and from there determine the curvature two-forms. This has been determined in some details in \cite{desitter2}, so here we will point out some of the finer points not discussed elsewhere. We start by first writing the curvature tensors in terms of the warped ({\it i.e.} $g_s$ dependent) metric components: 
\bg\label{curvten}
{\bf R}_{\rm IJKL} & = & \left({\bf g}_{\rm IL, JK} + {\bf g}_{\rm JK, IL}\right) - \left({\bf g}_{\rm IK, JL} + 
{\bf g}_{\rm JL, IK}\right) \nonumber\\
& + & {\bf g}^{\rm TS}\left({\bf g}_{\rm TI, K} + {\bf g}_{\rm TK, I} - {\bf g}_{\rm IK, T}\right) 
\left({\bf g}_{\rm JL, S} + {\bf g}_{\rm SL, J} - {\bf g}_{\rm SJ, L}\right) \nonumber\\
&-& {\bf g}^{\rm TS}\left({\bf g}_{\rm TJ, K} + {\bf g}_{\rm TK, J} - {\bf g}_{\rm JK, T}\right) 
\left({\bf g}_{\rm IL, S} + {\bf g}_{\rm SL, I} - {\bf g}_{\rm SI, L}\right) \nonumber\\ 
& + & \left({\bf g}_{\rm SI, K} + {\bf g}_{\rm SK, I} - {\bf g}_{\rm IK, S}\right) {\bf g}^{\rm PS}_{~~~,{\rm J}} 
{\bf g}_{\rm PL} 
 -  \left({\bf g}_{\rm SJ, K} + {\bf g}_{\rm SK, J} - {\bf g}_{\rm JK, S}\right) {\bf g}^{\rm PS}_{~~~,{\rm I}} 
{\bf g}_{\rm PL}. \nd
where $({\rm I, J, K, L})$ denote all the coordinates of the eight manifold. Using these components we would basically require to evaluate two curvature forms: ${\bf  R}^{a_o b_o}_{[ab]}$ and 
${\bf  R}^{a_o b_o}_{[{\rm MN}]}$. We start with the first one, whose explicit expression is given by:

{\footnotesize   
\bg\label{beachbb1}
{\bf  R}^{a_o b_o}_{ab}(y, g_s) &\equiv& {\bf R}_{ab{\rm M'N'}} ~e^{a_o {\rm M'}} e^{b_o {\rm N'}}  
+ {\bf R}_{aba'b'} ~e^{a_o a'} e^{b_o b'} + {\bf R}_{ab{\rm M'}0} ~e^{a_o {\rm M'}} e^{b_o 0} 
+ {\bf R}_{abij} ~e^{a_o i} e^{b_o j} \nonumber\\
&=&  g_s^2 ~{\rm R}^{a_o b_o}_{[ab]} + {\cal O}({\rm F}_1, {\rm F}_2) = \sum_{k \ge 0} {\rm R}^{(k)a_o b_o}_{[ab]}(y)
\left({g_s\over {\rm H}}\right)^{2 + 2k/3}, \nd}
where the explicit form for the Riemann tensors have been taken from {\bf Table \ref{firzas}} and 
$e^{a_o{\rm M}}$ denote the eleven-dimensional vielbeins. $({\rm M, N}) \in {\cal M}_4 \times {\cal M}_2$, $(a, b) \in {\mathbb{T}^2\over {\cal G}}$ and $(a_o, b_o)$ are the locally $SO(11)$ indices. Let us analyze the four Riemann tensors that enter in the definition of the curvature form \eqref{beachbb1}. The first one is ${\bf R}_{aba'b'}$ which could also be written as ${\bf R}_{abab}$. This takes the form:
\bg\label{bbeach1}
{\bf R}_{abab} = -\left({\bf g}_{aa, 0} {\bf g}_{bb, 0} - {\bf g}^2_{ab, 0}\right) {\bf g}^{00}  
-\left({\bf g}_{aa, {\rm M}} {\bf g}_{bb, {\rm N}} - {\bf g}_{ab, {\rm M}} {\bf g}_{ab, {\rm N}}\right) 
{\bf g}^{\rm MN}, \nd
where we see that even with {\it flat} toroidal direction the first term is non-zero because of it's $g_s$ dependence from \eqref{mup}. The third term tells us that we could go beyond flat metric and choose 
${\bf g}_{aa} = {\bf g}_{aa}(y, g_s)$. Finally,  
the second and the fourth terms require us to introduce a cross-term 
${\bf g}_{ab} \equiv {\bf g}_{ab}(y, g_s)$, that doesn't appear in \eqref{mup}. Where does it come from?

The answer is easy to see. The
existence of cross-term for the metric of ${\mathbb{T}^2\over {\cal G}}$, and the dependence of the metric on the coordinates $({\rm M, N}) \in {\cal M}_4 \times {\cal M}_2$ {\it i.e.} on the base of the eight-manifold appears from the constant coupling limit of F-theory \cite{DM} that we referred to earlier.  The
${\cal M}_2 \times {\mathbb{T}^2\over {\cal G}}$ manifold, which is locally a product geometry, goes to ${{\cal M}_4^{(2)} \over {\cal G}}$ globally with 
${\cal G} \equiv \mathbb{Z}_2, \mathbb{Z}_3, \mathbb{Z}_4$, and $\mathbb{Z}_6$ \cite{DM}. If we take the simplest case of 
${\cal G} = \mathbb{Z}_2$, it would imply a toroidal space $\mathbb{T}^2$ over all points of 
${\cal M}_2$ except at {\it four} points where the metric actually becomes an orbifold ${\mathbb{T}^2\over \mathbb{Z}_2}$. This means the choice of the unwarped metric component $g_{ab} = \delta_{ab}$ that we took in \eqref{mup}, holds at all points over ${\cal M}_2$ and only develops a cross-term $g_{ab}$ at four points. Near a fixed point, we could demand the metric to be:
\bg\label{siennamil}
{\bf g}_{ab}(y, g_s) = \left({g_s\over {\rm H}}\right)^{4/3} \left(\begin{matrix} 1 + f_1(y) & f_3(y) \cr f_3(y) & 1 + f_2(y)\end{matrix}\right), \nd 
where $f_i(y)$ are highly localized functions of $y = (y^m, y^\alpha)$.
Clearly then, we demand $g_{ab} \equiv 
g_{ab}(y^\alpha)$, or more generically, $g_{ab} \equiv g_{ab}(y)$, otherwise the duality map between IIB and M-theory will be ruined. Similar story can be elaborated for the next Riemann tensor 
${\bf R}_{{\rm MN}ab}$, which may be expressed as:
\bg\label{polameys}
{\bf R}_{{\rm MN}ab} = {\bf g}^{cd} \Big({\bf g}_{ac, {\rm M}} {\bf g}_{bd, {\rm N}} - {\bf g}_{ac, {\rm N}} 
{\bf g}_{bd, {\rm M}}\Big)
+ {\bf g}_{bd}\left({\bf g}_{ac, {\rm M}} {\bf g}^{cd}_{~~, {\rm N}} - 
{\bf g}_{ac, {\rm N}} {\bf g}^{cd}_{~~, {\rm M}}\right), \nd  
which requires us to keep both the diagonal components, ${\bf g}_{aa}(y, g_s)$ and ${\bf g}_{bb}(y, g_s)$, as well as off-diagonal components, ${\bf g}_{ab}(y, g_s)$, as non-trivial functions of $y \in {\cal M}_4 \times
{\cal M}_2$ and $g_s$. Our above explanation of these cross-components provide a nice realization of the present scenario. In fact this also explains the other two components appearing in \eqref{beachbb1}, namely,
${\bf R}_{ab{\rm M}0}$ and ${\bf R}_{abij}$, including the following curvature two-form:

{\footnotesize
\bg\label{beachbb2}
{\bf  R}^{a_o b_o}_{\rm MN}(y, g_s) &\equiv& {\bf R}_{\rm MNM'N'} ~e^{a_o {\rm M'}} e^{b_o {\rm N'}}  
+ {\bf R}_{{\rm MN}ab} ~e^{a_o a} e^{b_o b} + {\bf R}_{{\rm MNM'}0} ~e^{a_o {\rm M'}} e^{b_o 0} 
+ {\bf R}_{{\rm MN}ij} ~e^{a_o i} e^{b_o j} \nonumber\\
&=& g_s^0~ {\rm R}^{a_o b_o}_{[{\rm MN}]} + {\cal O}({\rm F}_1, {\rm F}_2) = \sum_{k \ge 0} 
{\rm R}^{(k)a_o b_o}_{[{\rm MN}]}(y)\left({g_s\over {\rm H}}\right)^{2k/3},  \nd}    
which tells us that to lowest order this is independent of $g_s$. The sum over $k$ for both \eqref{beachbb1} and \eqref{beachbb2} comes from the $({\rm F}_1, {\rm F}_2)$ factors, that include their products or 
even $\sqrt{{\rm F}_i}$, whose explicit $g_s$ expansion may be extracted from \eqref{saboot} in the limit 
$g_s << 1$. Combining everything together, the curvature two-form may be expressed in the following way:
\bg\label{beachmaa}
\mathbb{R} \equiv {\bf R}^{a_o b_o}_{\rm MN} {\bf M}_{a_o b_0} ~dy^{\rm M} \wedge dy^{\rm N} + 
{\bf R}^{a_o b_o}_{ab} {\bf M}_{a_o b_o} ~dy^a \wedge dy^b, \nd
where the holonomy matrices ${\bf M}_{a_o b_o}$ are chosen so as to facilitate the trace algebra, although 
they are not essential in the description of the curvature forms. In fact other definitions exist that do not involve the holonomy matrices. The constraint there is  to define traces properly. Here, introducing 
${\bf M}_{a_o b_o}$, makes this rather automatic because:
\bg\label{brianmaata}
&& {\rm tr}~{\bf M}_{a_o b_o} = 0, ~~ {\rm tr} \left({\bf M}_{a_o b_o} {\bf M}_{a_1 b_1}\right) \equiv 
\delta_{b_o a_1} ~\delta_{b_1 a_o} \nonumber\\
&& {\rm tr}\left({\bf M}_{a_o b_o}{\bf M}_{a_1 b_1}{\bf M}_{a_2 b_2}
{\bf M}_{a_3 b_3}\right) = \delta_{b_o a_1}~\delta_{b_1 a_2}~\delta_{b_2 a_3} ~\delta_{b_3 a_0}, 
\nd
thus facilitating the ensuing analysis. As an example, keeping $\mathbb{G} = 0$ in \eqref{kajalkir}, and then plugging the definition of $\mathbb{R}$ from \eqref{beachmaa} with the trace conditions from 
\eqref{brianmaata}, we can easily show that the ${\bf X}_8$ polynomial scales in the following way with respect to $g_s$: 

{\footnotesize
\bg\label{polameys}
{\bf X}_8(y, g_s) \equiv  {\rm tr}~ \mathbb{R}^4 - {1\over 4}\left({\rm tr}~\mathbb{R}^2\right)^2
=  g_s^2 \hat{\bf X}_8(y) + {\cal O}\left({\rm F}_1, {\rm F}_2\right)
= \sum_{k \ge 0} {\bf X}_8^{(k)}(y) \left({g_s \over {\rm H}}\right)^{2 + 2k/3}, \nd}
where $\hat{\bf X}_8(y) \equiv {\bf X}^{(0)}_8(y)$ is an eight-order form that only depends on the base ${\cal M}_4 \times {\cal M}_2$
of the eight-manifold. The independence of the toroidal direction ${\mathbb{T}^2\over {\cal G}}$ stems from the curvature tensors which are only dependent on the coordinates of the six-dimensional base. The integral of $\hat{\bf X}_8(y)$ over the eight-manifold is {\it not} necessarily related to the Euler characteristics of the eight-manifold because of the underlying non-K\"ahlerity of the base.  More importantly however is the 
dominant $g_s^2$ piece of ${\bf X}^{(0)}_8(y)$ telling us that the integral of ${\bf X}_8$ cannot be a 
time-{\it independent} quantity. In \cite{desitter2} we saw how this controls the anomaly cancellation conditions, but here we will see how this controls the time-dependences of the G-flux components.

Before moving ahead, let us ask whether we could make some parts of ${\bf X}_8$ independent of $g_s$. This could happen if we can make the dominant part of ${\bf R}^{a_ob_o}_{[ab]}$ $g_s$ independent as 
the dominant part of ${\bf R}^{a_ob_o}_{[{\rm MN}]}$ is already $g_s$ independent. This seems possible if we can make the toroidal metric components $y^b = x_{11}$ dependent, {\it i.e.}  
${\bf g}_{aa} = {\bf g}_{aa}(y, y^b, g_s), {\bf g}_{bb} = {\bf g}_{bb}(y, y^b, g_s)$ and 
${\bf g}_{ab} = {\bf g}_{ab}(y, y^b, g_s)$.  Such dependences will give additional contributions to 
${\bf R}_{abab}$ in \eqref{bbeach1} as:
\bg\label{bbeach3}
{\bf R}^{(+)}_{abab}(y, g_s) = {\bf g}_{aa, bb} - {\bf g}_{aa, b} {\bf g}_{ab} {\bf g}^{ab}_{~~, b} 
- {\bf g}_{aa, b} {\bf g}_{bb, b} {\bf g}^{bb} = {\rm R}^{(+)}_{abab}(y)\left({g_s\over {\rm H}}\right)^{4/3}, \nd
where no sum over repeated indices is implied. The $g_s$ scaling of such a term then implies that the two-form constructed out of it will scale as $g_s^0$. Plugging this in the definition of the ${\bf X}_8$ polynomial will at least provide a $g_s$ {\it independent} piece. Unfortunately this fails because of following three reasons.

\vskip.1in

\noindent $\bullet$ One, the inherent $y^b = x_{11}$ dependence cannot be motivated from the type IIB metric \eqref{iibmet} and in fact the 
$x_{11}$ dependence of ${\bf g}_{33}$ will imply that somehow the IIB metric {\it knows} about the eleven-dimensional uplift. This is also one of the reasons for not taking the localized form $\Omega_{ab}$, that quantifies the localized G-flux components ${\bf G}_{{\rm MN}ab}$, to be $x_{11}$ dependent but only 
$(y^m, y^\alpha)$ dependent in \cite{desitter3, coherbeta}. 

\vskip.1in

\noindent $\bullet$ Two, in the $g_s$ scaling \eqref{ddeluca} for the perturbative quantum terms \eqref{phingsha3}
we have to keep $n_3 = 0$, where $n_3$ is the number of derivatives along $(x_3, x_{11})$ directions, so as to avoid conflicting with the $g_s$ hierarchy. In fact taking $k'_3 \ge {3\over 2}$ in 
\eqref{ddeluca}, as mentioned therein, is basically to restore the $g_s$ hierarchy. 

\vskip.1in

\noindent $\bullet$ Three, all the Riemann curvature components contribute as $+{2l_i\over 3}$ to the $g_s$ scaling in \eqref{ddeluca} because of their specific $g_s$ expansions given in {\bf Table \ref{firzas}}. However if we now demand that the dominant $g_s$ scaling of ${\bf R}_{abab}$ is $g_s^{4/3}$ instead of $g_s^{10/3}$, then it would contribute as $-{4l_2\over 3}$ to \eqref{ddeluca}. This relative minus sign would be a problem if we want to retain the $g_s$ hierarchy, and in fact there is no way we could revert the sign here\footnote{For G-flux components any relative minus sign can be reverted back to positive by switching on appropriate $k'_i$ in \eqref{ddeluca}. This cannot happen for the curvature terms. Also, other curvature components could start having  different $g_s$ scaling than the ones shown in {\bf Table \ref{firzas}}, thus creating more relative minus signs in 
\eqref{ddeluca}.}, signalling a stronger breakdown of four-dimensional EFT description. 

\vskip.1in

All in all, it appears that making the toroidal metric components dependent on $x_{11}$ direction {\it cannot} provide a consistent picture with $g_s$ independent ${\bf X}_8$ piece. 

The above discussion points to the fact that the $g_s$ independent ${\bf X}_8$ polynomial defined over the internal eight-manifold would in principle clash with a four-dimensional EFT description (due to the loss of $g_s$ and subsequently the ${\rm M}_p$ hierarchies). We could also try to study the behavior of other 
${\bf X}_8$ polynomials that are not restricted over the internal eight-manifold. These polynomials, if they exist, would have to be defined over certain odd-dimensional cycles inside the eight-manifold. They would also have to depend on the behavior of the curvature two-forms. For us we would require the following curvature forms, in addition to what we had earlier:
\bg\label{polamaala2}
{\bf  R}^{a_o b_o}_{0b} &=&  {\bf R}_{0b0a} ~e^{a_o 0} e^{b_o a} = \sum_{k \ge 0} 
{\rm R}^{(k)a_o b_o}_{[ab]} \left({g_s\over {\rm H}}\right)^{2k/3} \nonumber\\
{\bf  R}^{a_o b_o}_{0{\rm M}} 
&\equiv& {\bf R}_{0{\rm MPQ}} ~e^{a_o {\rm P}} e^{b_o {\rm Q}}  
+ {\bf R}_{0{\rm M}ij} ~e^{a_o i} e^{b_o j} + {\bf R}_{0{\rm M}ab} ~e^{a_o a} e^{b_o b} 
+ {\bf R}_{0{\rm M}0{\rm N}} ~e^{a_o 0} e^{b_o {\rm N}}\nonumber\\
&=&  g_s^{-1} ~{\rm R}^{a_o b_o}_{[0{\rm M}]} + {\cal O}({\rm F}_1, {\rm F}_2) = 
\sum_{k \ge 0} {\rm R}^{(k)a_ob_o}_{[0{\rm M}]}(y) \left({g_s\over {\rm H}}\right)^{2k/3 - 1}
\nonumber\\
{\bf  R}^{a_o b_o}_{ij} 
&\equiv& {\bf R}_{ij{\rm PQ}} ~e^{a_o {\rm P}} e^{b_o {\rm Q}}  
+ {\bf R}_{iji'j'} ~e^{a_o i'} e^{b_o j'} + {\bf R}_{ijab} ~e^{a_o a} e^{b_o b} 
+ {\bf R}_{ij0{\rm N}} ~e^{a_o 0} e^{b_o {\rm N}} \nonumber\\
&=&  g_s^{-2} ~{\rm R}^{a_o b_o}_{[ij]} + {\cal O}({\rm F}_1, {\rm F}_2) = 
\sum_{k \ge 0} {\rm R}^{(k)a_o b_o}_{[ij]}(y) \left({g_s\over {\rm H}}\right)^{2k/3 - 2}, 
\nd
where note that the first curvature two-form do have the usual $g_s$ expansion despite the vielbeins not having any  ${\rm F}_i$ dependences, because the curvature tensors themselves have hidden ${\rm F}_i$ 
dependences (see {\bf Table \ref{firzas}}). The other curvature two-forms are expanded in powers of $g_s$ as shown above, although note that the dominant $g_s$ scalings are $-1$ and $-2$ respectively. We can combine \eqref{polamaala2}, \eqref{beachbb1} and \eqref{beachbb2} to construct the following generic curvature two-form:

{\footnotesize
\bg\label{beachmaaa}
\mathbb{R}_{\rm tot} \equiv \mathbb{R} + {\bf R}^{a_o b_o}_{0{\rm M}} {\bf M}_{a_o b_0} ~dx^0 \wedge 
dy^{\rm M} + 
{\bf R}^{a_o b_o}_{ij} {\bf M}_{a_o b_o} ~dx^i \wedge dx^j + {\bf R}^{a_o b_o}_{0b} {\bf M}_{a_o b_0} ~dx^0 \wedge dy^b, \nd}
where now the dependence goes beyond the internal eight-manifold to encompass the full eleven-dimensional manifold. We have also defined $\mathbb{R}$ as the curvature form derived earlier in
\eqref{beachmaa}. We can now use this definition of the curvature two-form to construct three other polynomials by replacing $\mathbb{R}$ in the definition of ${\bf X}_8$
 by $\mathbb{R}_{\rm tot}$. The first of these three polynomials can be defined over the eight-manifold in the following way:  
\bg\label{polathai1}
{\bf X}_8({y}, g_s)\Big\vert_{\mathbb{C}^{(1)}} = g_s^{-3} {\bf X}^{(1)}_8({y}) + {\cal O}({\rm F}_1, {\rm F}_2) = 
\sum_{k \ge 0} {\bf X}_8^{(k, 1)}({y}) \left({g_s\over {\rm H}}\right)^{2k/3 - 3}, \nd
where $\mathbb{C}^{(1)} \equiv  {\bf R}^{2, 1} \times \mathbb{C}_5$ with $\mathbb{C}_5$ being a  five cycle in ${\cal M}_4 \times {\cal M}_2$. Note the dominant $g_s$ scaling for such a polynomial is $-3$. In a similar vein we define the second of the three polynomials in the following way:
\bg\label{polathai2}
{\bf X}_8({y}, g_s)\Big\vert_{\mathbb{C}^{(2)}} = g_s^{-2} {\bf X}^{(2)}_8({y}) + {\cal O}({\rm F}_1, {\rm F}_2) = 
\sum_{k \ge 0} {\bf X}_8^{(k, 2)}({y}) \left({g_s\over {\rm H}}\right)^{2k/3 - 2}, \nd
now defined over the eight-manifold in a similar way as in \eqref{polathai1}, {\it i.e.} 
$\mathbb{C}^{(2)} \equiv  {\bf R}^{2, 1} \times \mathbb{C}_4 \times {\bf S}^1$ with  
$\mathbb{C}_4 \in {\cal M}_4 \times {\cal M}_2$ and the one-cycle ${\bf S}^1 \in {\mathbb{T}^2 \over {\cal G}}$. The dominant $g_s$ scaling of the polynomial is $-2$ so behaves differently from \eqref{polathai1}.
Finally, the third of the three polynomials take the following form:
\bg\label{polathai3}
{\bf X}_8({y}, g_s)\Big\vert_{\mathbb{C}^{(3)}} = g_s^{-1} {\bf X}^{(3)}_8({y}) + {\cal O}({\rm F}_1, {\rm F}_2) = 
\sum_{k \ge 0} {\bf X}_8^{(k, 3)}({y}) \left({g_s\over {\rm H}}\right)^{2k/3 - 1}, \nd
where $\mathbb{C}^{(2)} \equiv {\bf R}^{2, 1} \times \mathbb{C}_3 \times {\mathbb{T}^2\over {\cal G}}$ with 
$\mathbb{C}_3 \in {\cal M}_4 \times {\cal M}_2$, {\it i.e.} a three-cycle on the base. Such a polynomial has a dominant $g_s$ scaling of $-1$ which should be compared to the dominant $g_s$ scalings of the other two polynomials in \eqref{polathai1} and \eqref{polathai2}. All the three polynomials are also defined over 
{\it odd} cycles in the internal manifold, so globally there is a possibility that 
none of these odd cycles might exist. However these manifolds are also non-K\"ahler, so we cannot rule out odd-cycles now\footnote{More so, the internal manifold takes the form ${\cal M}_4 \times {\cal M}_2 \times 
{\mathbb{T}^2\over {\cal G}}$ only locally, and therefore one expects more complicated global topology.}. In addition to that, these polynomial forms appear in the flux EOMs (see \cite{desitter2}) which only care about local terms, so existence or non-existence of global odd-cycles will not pose any issues for at least the flux EOMs. Of course global constraints matter, but since all the three eight-manifolds 
$\mathbb{C}^{(i)}$ are non-compact $-$ and including the fact that the internal manifold is by itself 
non-K\"ahler $-$ global constraints will not pose any restrictions here. With these at hand, let us study the flux EOMs.


\subsection{Analysis of the flux equations and time-dependent background \label{flox}}

Let us go back to the M-theory action \eqref{sheela2}, and express it in a way from where the EOMs for the G-flux components may be easily derived. First however note the following re-definition:
\bg\label{aliceL3}
\int {\bf G}_4 \wedge \ast_{11} \mathbb{Y}_4& \equiv & \int d^{11} y \sqrt{-{\bf g}_{11}} 
\sum_{\{l_i\}, n_i}
\mathbb{Q}_{\rm T}^{\{l_i\}, n_i}\\
&=& \int d^{11} y \sqrt{-{\bf g}_{11}} 
\left({\bf G}_4\right)_{{\rm M_1M_2M_3M_4}} \left(\mathbb{Y}_4\right)_{{\rm N_1N_2N_3N_4}} 
{\bf g}^{\rm M_1N_1}
{\bf g}^{\rm M_2N_2}{\bf g}^{\rm M_3N_3}{\bf g}^{\rm M_4N_4},\nonumber \nd
which may be used as the defining relation for the four-form $\mathbb{Y}_4$. Thus $\mathbb{Y}_4$ contains 
{\it all} possible perturbative interactions contained in the sum over all set of  $\{l_i\}$ and $n_i$ in 
$\mathbb{Q}_{\rm T}^{\{l_i\}, n_i}$ from \eqref{phingsha3}. Using \eqref{aliceL3}, 
we can express the perturbative parts of the action \eqref{sheela2} in the following compact way:
{\footnotesize
\bg\label{weisz2}
{\bf S}_{\rm pert} \equiv b_1\int {\bf G}_4 \wedge \ast_{11} {\bf G}_4 + b_2 \int {\bf C}_3 \wedge {\bf G}_4
\wedge {\bf G}_4 + b_3 \int {\bf C}_3 \wedge \mathbb{Y}_8 + b_4 \int {\bf G}_4 \wedge \ast_{11}
\mathbb{Y}_4 + n_b \int {\bf C}_3 \wedge {\bf \Lambda}_8, \nonumber\\ \nd}
where $b_i$ are typically proportional to powers of ${\rm M}_p$ that may be extracted from \eqref{sheela2}; and $\mathbb{Y}_8$ is as given in \eqref{kajalkir}. $\bf{\Lambda}_8$ is a localized eight-form that fixes the positions of $n_b$ number of M2-branes (with an equivalent term for $\bar{n}_b$ number of $\overline{\rm M2}$-branes). There are also non-perturbative terms appearing from \eqref{sheela2} contributing to the flux Lagrangian, and we will discuss them later after we study the consequence of the perturbative quantum terms on the flux EOMs. We will however not study all the flux EOMs here, but concentrate only on 
${\bf G}_{0ij{\rm M}}$ and ${\bf G}_{{\rm MNP}a}$ EOMs. The remaining EOMs may be studied from 
\cite{desitter2}. 

The EOM for the G-flux components ${\bf G}_{0ij{\rm M}}$ is interesting in its own right. In \cite{DRS} and 
\cite{1402}, this EOM was responsible to determine the consistency condition, {\it i.e.} the Gauss' law constraint, for the system. For the present case, it takes the following form:
\bg\label{evaB100}
&& b_1 \partial_{N_8}\Big(\sqrt{-{\bf g}_{11}}~{\bf G}_{0ij{\rm M}} ~{\bf g}^{00'} {\bf g}^{ii'} {\bf g}^{jj'} {\bf g}^{\rm MM'}\Big)
\epsilon_{0'i'j'{\rm M'N_1 N_2 ...... N_7}}\\
&& ~~~= b_4~\partial_{\rm N_8}\Big(\sqrt{-{\bf g}_{11}} \left(\mathbb{Y}_4\right)_{0ij{\rm M}} 
{\bf g}^{00'} {\bf g}^{ii'} {\bf g}^{jj'} {\bf g}^{\rm MM'}\Big)\epsilon_{0'i'j'{\rm M'N_1.....N_7}}\nonumber\\
&&~~~+ b_2~{\bf G}_{\rm N_1.....N_4} {\bf G}_{\rm N_5.....N_8} 
+ b_3 \left(\mathbb{Y}_8\right)_{\rm N_1......N_8} + 
{\rm T}_2\Big(n_b \delta^8(y - y_1) - \bar{n}_b \delta^8(y - y_2)\Big) \epsilon_{\rm N_1.....N_8}, \nonumber\nd
where $\epsilon_{0'...{\rm N}_7}$ is tensor density; and 
which on face value appears to be similar to what we had in \cite{DRS} and \cite{1402}. However once we plug-in the $g_s$ dependences of the G-flux components from {\bf Table \ref{firzas2}}, one can start seeing the differences.  Expanding in powers of $g_s$, both from the fluxes and the ${\rm F}_i(t)$ factors in 
\eqref{saboot}, \eqref{evaB100} changes to:

{\footnotesize
\bg\label{evaB1111}
&&-\square {\rm H}^4\sum_{\{k_i\}} b_1{\rm C}^{(1)}_{k_1} {\rm C}^{(2)}_{k_2}\left({g_s\over {\rm H}}\right)^{2(k_1 + k_2)/3}
+ {b_1\over \sqrt{g_8}}\sum_{\{k_3, k'_i\}} \partial_{\rm N_8}\Big(\sqrt{g_8} ~{\rm H}^8 {\cal G}^{(k_3)}_{012{\rm M}} g^{\rm MN_8}\Big) 
{\rm C}^{(1)}_{k'_1} {\rm C}^{(2)}_{k'_2} \left({g_s\over {\rm H}}\right)^{2(k_3 + k'_1 + k'_2)/3} \nonumber\\
&& ~~~~= b_2 \sum_{\{k_i\}} {\cal G}^{(k_4)}_{\rm N_1... N_4} \left(\ast_8{\cal G}^{(k_5)}\right)^{\rm N_1...N_4}
\left({g_s\over {\rm H}}\right)^{2(k_4 + k_5)/3} + {b_3 \over \sqrt{g_8}} \sum_{k_6 \ge 0}\left({\bf X}^{(k_6)}_8\right)_{\rm N_1...N_8} \epsilon^{\rm N_1...N_8}\left({g_s\over {\rm H}}\right)^{2 + 2k_6/3} \nonumber\\
&& ~~~~+ {b_4\over \sqrt{g_8}}\sum_{\{k\}} \partial_{\rm N_8}\Big(\sqrt{g_8}\left(\mathbb{Y}^{(k)}_4\right)^{012N_8}\Big)
\left({g_s\over {\rm H}}\right)^{\theta_k - 2/3} +  {{\rm T}_2\over \sqrt{g_8}}
\Big(n_b \delta^8(y - y_1) - \bar{n}_b \delta^8(y - y_2)\Big), 
\nd}
where $\theta_k \equiv \theta_{lkk'\hat{k}}$ from \eqref{ddeluca}; $\square$, $\ast_8$ and $g_8$ are all defined with respect to the un-warped metric of the internal eight-manifold;
${\rm C}^{(i)}_{k_i}$ are the coefficients appearing in \eqref{saboot}, ${\rm T}_2$ is the M2 or $\overline{\rm M2}$-brane tension, the branes being localized at 
$y_1$ and $y_2$ points on the internal sub-manifold ${\cal M}_4 \times {\cal M}_2$ respectively.  Multipole M2 or 
$\overline{\rm M2}$-branes on top of each other should have more non-trivial action compared to what appears in \eqref{sheela2} or \eqref{weisz2}, maybe along the lines of \cite{mukhi}, but we will not indulge in these details here as they don't effect our analysis. We also expect:
\bg\label{hurdwood}
k_1 + k_2 = k_3 + k'_1 + k'_2 = k_4 + k_5 = 2 + {k_6\over 3} = \theta_k - {2\over 3}, \nd
at least  for $k_i > 0$ and $k'_i > 0$. We have also taken $k_3 > 0$ in \eqref{evaB1111}, but the remaining $k_i \ge 0$. In particular, we want to see whether $k_4 = k_5 = 0$ {\it i.e.} whether they can ever vanish, as the vanishing of $k_4$ and $k_5$ would imply the possibility of time-independent parts of the G-flux components ${\bf G}_{{\rm MNP}a}$ and ${\bf G}_{\rm MNPQ}$. Clearly $(k_1, k_2, k'_1, k'_2, k_6)$ can vanish but, as we mentioned above, $k_3 > 0$. Thus if some components of G-flux become time-independent, it cannot get support from the second and the ${\bf X}_8^{(k_6)}$ terms of \eqref{evaB1111}. Out of the remaining terms, that could support time-independent G-flux components, the quantum contributions, i.e the $\mathbb{Y}_4^{(k)}$ pieces, appear as {\it total derivatives} in \eqref{evaB1111} and therefore they cannot contribute globally.  The {\it local} contributions then give us:
\bg\label{rebhall}
-\square {\rm H}^4 &= &{b_2\over b_1}~{\cal G}^{(0)}_{\rm N_1... N_4} 
\left(\ast_8{\cal G}^{(0)}\right)^{\rm N_1...N_4} 
+ {b_4\over b_1 \sqrt{g_8}}\sum_{\{k\}} \partial_{\rm N_8}\Big(\sqrt{g_8}\left(\mathbb{Y}^{(k)}_4\right)^{012N_8}\Big)\delta\left(\theta_k - {2\over 3}\right) \nonumber\\
 &&~~~~~~~~~ + {{\rm T}_2\over b_1 \sqrt{g_8}}
\Big(n_b \delta^8(y - y_1) - \bar{n}_b \delta^8(y - y_2)\Big),   \nd
where the delta-function localizes $\theta_k \equiv \theta_{lkk'\hat{k}} = {2\over 3}$ in \eqref{saboot}. Looking at the form of $\mathbb{Y}^{012{\rm N_8}}_4$ from \eqref{aliceL3}, its coupling to 
${\bf G}_{012{\rm N_8}}$, and the fact that the $g_s$ scaling is only ${2\over 3}$, would imply that 
it can only be ${\cal G}^{(0)012{\rm N_8}}$. But this is exactly the first term $\square {\rm H}^4$, so the quantum term here does nothing but renormalize the coefficient of $\square {\rm H}^4$. This unfortunately leaves only the first and the third term from \eqref{rebhall}, {\it i.e.}:
\bg\label{arkansas}
-\square {\rm H}^4 = {b_2\over \hat{b}_1}~{\cal G}^{(0)}_{\rm N_1... N_4} 
\left(\ast_8{\cal G}^{(0)}\right)^{\rm N_1...N_4} + {{\rm T}_2\over \hat{b}_1 \sqrt{g_8}}
\Big(n_b \delta^8(y - y_1) - \bar{n}_b \delta^8(y - y_2)\Big), \nd
where $\hat{b}_1$ is the renormalized constant (note: since it appears equally on all terms, the precise renormalized value doesn't matter much). The equation \eqref{arkansas} could in principle serve as the key equation for determining the warp-factor ${\rm H}(y)$ provided we can argue the existence of time-{\it independent} G-flux components. To see this, we multiply both sides of \eqref{arkansas} with un-warped
$\sqrt{g_8}$, and then integrate over the eight-manifold to get the following consistency condition:
\bg\label{palstodio}
b_2 \int d^8 y \sqrt{g_8} ~{\cal G}^{(0)}_{\rm N_1... N_4} \left(\ast_8{\cal G}^{(0)}\right)^{\rm N_1...N_4} 
+ {\rm T}_2 \left(n_b - \bar{n}_b\right) = 0, \nd
where ${\cal G}^{(0)}_{\rm N_1... N_4}$ only involve those G-flux components that we keep time-independent, {\it i.e.} ${\cal G}^{(0)}_{{\rm MNP}a}$ and (possibly) ${\cal G}^{(0)}_{\rm MNPQ}$.  
The condition \eqref{palstodio} doesn't quite dis-allow time-independent components if $\bar{n}_b > n_b$, i.e if the number of $\overline{\rm M2}$-branes are bigger than the number of M2-branes. Note that, the presence of $\overline{\rm M2}$-branes (or even M2-branes) are not strictly required in our analysis as the supersymmetry is broken by non self-dual G-fluxes. Unfortunately presence of multiple $\overline{\rm M2}$ is problematic in its own right as has been pointed out by \cite{bena}, implying that taking $\bar{n}_b = 0$ or 
$\bar{n}_b = 1$ should still be considered safe. If $n_b \ge 1$, then this will put a very tight constraint on the time-independent G-flux components, and there appears a good chance that, with quantized fluxes, we may not have a solution here. 

The G-flux components ${\bf G}_{{\rm MNP}a}$ are related to the three-form flux components, ${\bf F}_3$ and ${\bf H}_3$, in the type IIB side. 
The situation gets trickier once we look at the EOM of these components in M-theory. The ${\bf G}_{{\rm MNP}a}$
EOM takes the following form:
\bg\label{evaoo7}
&& b_1~ \partial_{\rm N_8}\Big(\sqrt{-{\bf g}_{11}}~{\bf G}_{{\rm MNP}a} ~{\bf g}^{\rm MM'} {\bf g}^{\rm NN'} 
{\bf g}^{\rm PP'} 
{\bf g}^{aa'}\Big)
\epsilon_{{\rm M'N'P'}a'{\rm N_1 N_2 ...... N_7}}\nonumber\\
&&~~~= b_2~{\bf G}_{0ij{\rm M}} {\bf G}_{{\rm QRN}b}~\delta^{[0}_{[{\rm N_1}} \delta^i_{\rm N_2} \delta^j_{\rm N_3}\delta^{\rm M}_{\rm N_4}
\delta^{\rm Q}_{\rm N_5} \delta^{\rm R}_{\rm N_6} \delta^{\rm N}_{\rm N_7} \delta^{b]}_{{\rm N_8}]} 
+ b_3 \left({\bf X}_8\right)_{\rm N_1......N_8}\nonumber\\
&& ~~~+ b_4~\partial_{\rm N_8}\Big(\sqrt{-{\bf g}_{11}} \left(\mathbb{Y}_4\right)_{{\rm MNP}a} 
{\bf g}^{\rm MM'} {\bf g}^{\rm NN'} {\bf g}^{\rm PP'} {\bf g}^{aa'}\Big)\epsilon_{{\rm M'N'P'}a'{\rm N_1.....N_7}}, \nd 
where the eight-form ${\bf X}_8$ should now be picked up from \eqref{polathai2}, provided of course that the 
cycle $\mathbb{C}_2$ exists. The quantum terms $\mathbb{Y}_4$ should again be read from \eqref{aliceL3}, as we did earlier, and its specific component will guide the precise form of it. All of these have been discussed earlier in great details in \cite{desitter2}, so won't go on further elaboration of this. Putting everything together, the component form of the above EOM becomes:

{\footnotesize
\bg\label{evaooo7}
&&b_1 \sum_{\{k_i\}}\partial_{\rm N_8}\Big(\sqrt{{g}_{8}}~{\cal G}^{(k_1)}_{{\rm MNP}a} ~{g}^{\rm MM'} 
{g}^{\rm NN'} {g}^{\rm PP'} 
{g}^{aa'}\Big) \left({g_s\over {\rm H}}\right)^{2k_1/3 -4}\epsilon_{{\rm M'N'P'}a'{\rm N_1 N_2 ...... N_7}}\nonumber\\
&&~~~ \times \left(a_{(mn\alpha)} + \left(a_{(mnp)}~{\rm C}^{(1)}_{k_2} {\rm C}^{(1)}_{k_3} {\rm C}^{(2)}_{k_4}  + a_{(m\alpha\beta)}~
{\rm C}^{(2)}_{k_2} {\rm C}^{(2)}_{k_3} {\rm C}^{(2)}_{k_4}\right) \left({g_s\over {\rm H}}\right)^{2(k_2 + k_3 + k_4)/3}\right)\nonumber\\
&&~~~= b_2\sum_{\{k_i\}}{\cal G}^{(k_5)}_{0ij{\rm M}} {\cal G}^{(k_6)}_{{\rm QRN}b}~
\left({g_s\over {\rm H}}\right)^{2(k_5 + k_6)/3 - 4}
\delta^{[0}_{[{\rm N_1}} \delta^i_{\rm N_2} \delta^j_{\rm N_3}\delta^{\rm M}_{\rm N_4}
\delta^{\rm Q}_{\rm N_5} \delta^{\rm R}_{\rm N_6} \delta^{\rm N}_{\rm N_7} \delta^{b]}_{{\rm N_8}]}
\nonumber\\ 
&& ~~~+ b_4 \sum_{\{k,k'\}}\partial_{\rm N_8}\Big(\sqrt{{g}_{8}} (\mathbb{Y}^{(k)}_4)^{{\rm M'N'P'}a'} 
\Big)\left({g_s\over {\rm H}}\right)^{\theta_k - 2k/3 - 14/3} \epsilon_{{\rm M'N'P'}a'{\rm N_1.....N_7}}
\nonumber\\ 
&& ~~~+ b_3 \sum_{k_7 \ge 0} \left({\bf X}^{(k_7, 2)}_8\right)_{N_1......N_8}\left({g_s \over {\rm H}}\right)^{2k_7/3 - 2}, \nd}
where the terms with coefficients $b_1$ and $b_4$ appear as total derivatives. The coefficients, for example 
$a_{(mn\alpha)}$, are constants and they simply allow us to selectively choose how the G-flux components arrange their legs along the six-dimensional base ${\cal M}_4 \times {\cal M}_2$.  The ${\bf X}_8^{(k_7, 2)}$ are time-independent components of ${\bf X}_8$ as defined in \eqref{polathai2}; and $\theta_k \equiv 
\theta_{lkk'\hat{k}}$ as in \eqref{ddeluca}. It is clear that, for $k_i > 0$ and $k > 0$, we expect:
\bg\label{bindiyag}
{2\over 3}(k_1 + k_2 + k_3 + k_4) - 4 = {2\over 3}(k_5 + k_6) - 4 = {2k_7\over 3} - 2 = \theta_k - {2k\over 3} 
- {14\over 3}, \nd
at least when we take $a_{(mnp)}$ or $a_{(m\alpha\beta)}$, whereas if we take $a_{(mn\alpha)}$, the first equality will not have $(k_2, k_3, k_4)$ pieces.

The aim of the above exercise is not to solve the EOM \eqref{evaooo7} exactly but to figure out whether 
$k_1 = k_6 = 0$. If this is the case then at least the components ${\bf G}_{{\rm MNP}a}$ will remain time-independent as before. Putting $k_1 = 0$, the first term scales as $g_s^{-4}$, as $(k_2, k_3, k_4) \ge 0$. The second term, i.e the one with $b_2$ coefficient, also scales as $g_s^{-4}$ as $k_5 \ge 0$. 
The ${\bf X}^{(0, 2)}_8$ scales as $g_s^{-2}$. The quantum terms now have two possible scalings, given by:
\bg\label{rakheeS}
\theta_k \equiv \theta_{lkk'\hat{k}} = {2\over 3}(k + 1), ~~~~~
\theta_k \equiv \theta_{lkk'\hat{k}} = {2\over 3}(k + 4), \nd
depending on how we want to balance the terms of the EOM. If $k = 0$, then the first choice in 
\eqref{rakheeS} will not give anything significant whereas the second choice, again for $k = 0$, will allow quartic order quantum terms. Since the terms with coefficients $b_2$ and $b_3$ scale differently when 
$k_i = 0$, and the terms with coefficients $b_1$ and $b_4$ appear with total derivatives, the integral constraint for the flux components become:
\bg\label{shakalmey}
\int d^8 y ~{\cal G}^{(0)}_{0ij{\rm M}} {\cal G}^{(0)}_{{\rm QRN}b} = 0, \nd
which unfortunately puts even a stronger constraint than what we encountered in \eqref{palstodio} because 
${\cal G}^{(0)}_{0ij{\rm M}} \propto \epsilon_{0ij{\rm M}}$. Putting everything together it appears, at least in the  presence of the perturbative corrections, that the time-independent parts of the G-flux components ${\bf G}_{{\rm MNP}a}$, if they exist, cannot be non-trivial.

What happens once we incorporate non-perturbative corrections to the two G-flux  EOMs discussed above, namely, \eqref{evaB100} and \eqref{evaoo7}; or more appropriately, to the EOMs in terms of their $g_s$ expansions in \eqref{evaB1111} and \eqref{evaooo7}? These non-perturbative terms should be related to the BBS \cite{BBS} and KKLT \cite{KKLT} type instanton gases, and have recently been discussed in 
\cite{desitter2, coherbeta}. Here we want to see how they could influence the G-flux EOMs discussed above by first expressing the integral of the exponential piece, as it appears in the last line of \eqref{sheela2}, in the following way:

{\footnotesize
\bg\label{nikolK}
{\bf S}^{(\{l_i\}, n_i)}_6(x) & = & \int d^6 y \sqrt{{\bf g}_6(y, g_s)} ~\mathbb{F}^{(1)}(x - y)  \mathbb{V}_2(y, g_s)
\mathbb{Q}_{\rm T}^{(\{l_i\}, n_i)}(y, g_s)\\
& = &  \int d^6 y \sqrt{{\bf g}_6(y, g_s)} ~\mathbb{F}^{(1)}(x - y)  \mathbb{V}_2(y, g_s)
\left({\bf G}_4\right)_{{\rm M_1M_2M_3M_4}} \left(\mathbb{Y}^{(\{l_i\}, n_i)}_4\right)_{{\rm N_1N_2N_3N_4}}
\prod_{i = 1}^4 {\bf g}^{{\rm M}_i {\rm N}_i},\nonumber  \nd} 
where $\mathbb{F}^{(1)}(x - y)$ is the non-locality function. We can in fact extract a {\it local} piece out of 
\eqref{nikolK} by expressing $\mathbb{F}^{(1)}(x - y) = \mathbb{F}^{(1)}(- y) 
+ {\cal O}\left(x {\partial \mathbb{F}^{(1)}(- y) \over \partial y}\right)$ in the limit the growth of 
$\mathbb{F}^{(1)}(-y)$ is smaller than any scale in the theory. However this decomposition is not necessary because, to identify with the BBS \cite{BBS} or the KKLT \cite{KKLT} type instanton gases, one needs to take into account the full function 
$\mathbb{F}^{(1)}(x-y)$ as well as  the orientations and the two-volume 
$\mathbb{V}_2(y, g_s)$ (see section 3.2 of \cite{coherbeta}). In general, we can use \eqref{nikolK} to express the action as:
\bg\label{swoodli}
{\bf S}_{\rm inst} = \sum_{\{l_i\}, n_i} \int d^{11}x \sqrt{-{\bf g}_{11}(x)}~ \sum_{k\ge 1} c_k 
\left[{\rm exp}\left(-k {\rm M}_p^{6-\sigma_i} {\bf S}_6^{(\{l_i\}, n_i)}(x)\right) - 1\right], \nd
where $c_k$ is a constant, and $\sigma_i \equiv \sigma(\{l_i\}, n_i)$ is the ${\rm M}_p$ dimension of 
$\mathbb{Q}_{\rm T}^{(\{l_i\}, n_i)}$. The presence of $-1$ piece has been explained in \cite{coherbeta}, and is important to remove some of the unnecessary terms that appear in the energy-momentum tensor. For the flux EOMs this is a redundant factor. Using \eqref{swoodli} we can easily
extract the contributions of the BBS or the KKLT type instantons to the two flux EOMs discussed above. The first contribution becomes:
\bg\label{sarbutler}
{\delta {\bf S}_{\rm inst}\over \delta {\bf C}_{012}(z, g_s)} &= & -\sum_{\{l_i\}, n_i}  \int d^{11}x \sqrt{-{\bf g}_{11}(x)}~
\sum_{k\ge 1} k c_k~{\rm exp}\left(-k {\rm M}_p^{6-\sigma_i} {\bf S}_6^{(\{l_i\}, n_i)}(x)\right)\nonumber\\
&&~~~~~ \times 
{\partial \over \partial z^{\rm M}}\Bigg(\sqrt{{\bf g}_6(z, g_s)} \mathbb{F}^{(1)}(x - z) \mathbb{V}_2(z, g_s)
 \mathbb{Y}_4^{(\{l_i, n_i\})012{\rm M}}(z, g_s)\Bigg),\nd
where we will assume that $\mathbb{F}^{(1)}(x - z)$ has no $g_s$ or ${\rm M}_p$ scalings (which is the simplest case considered in \cite{desitter3}). The action ${\bf S}_6^{(\{l_i\}, n_i)}(x)$ in \eqref{nikolK} scales as $g_s^{\theta_k - 2/3}$ where $\theta_k = \theta_{lkk'\hat{k}}$ in \eqref{phingsha3}. On the other hand, 
\eqref{sarbutler} scales in the following way:
\bg\label{sarbut2}
{\delta {\bf S}_{\rm inst}\over \delta {\bf C}_{012}(z)} \propto \sum_{ q\ge 1} q c_q~g_s^{\theta_{k} - 4/3} ~
{\rm exp}\left(-q g_s^{\theta_k - 2/3}\right), \nd
where for simplicity we hide the other factors (which will be elaborated on below) and $k$ is different from the $q$ that appears in the summation as should be clear from the context. The terms on the RHS of 
\eqref{sarbutler} or \eqref{sarbut2} should contribute to the G-flux EOM \eqref{evaB1111}. To see whether there could be a time-independent contribution from \eqref{sarbutler} or \eqref{sarbut2}, we collect all terms from $\mathbb{Y}_4^{(\{l_i, n_i\})012{\rm M}}(z, g_s)$ that scale as:
\bg\label{siegner}
\theta_k \equiv \theta_{lkk'\hat{k}} = {4\over 3}, \nd
in \eqref{phingsha3} and in \eqref{ddeluca}. Plugging \eqref{siegner} to the exponential piece in \eqref{sarbut2} tells us that it scales as ${\rm exp}\left(-q g_s^{2/3}\right)$ which, in the limit $g_s \to 0$, can have a perturbative expansion. This is in fact made possible because we took $\mathbb{F}^{(1)}(x - z)$ to have no $g_s$ scaling (taking the simplest case from \cite{desitter3}). Expanding the exponential piece perturbatively, and using the following 
integral condition:
\bg\label{ckeaton}
\int d^{11} x \sqrt{-g_{11}(x)} ~\mathbb{F}^{(1)}(x - z) \equiv \mathbb{F}(z), \nd
where $g_{11}(x)$ is the $g_s$ independent part of the determinant of the eleven-dimensional metric and 
$\mathbb{F}(z)$ is a local function. $\mathbb{F}(z)$ is also a finite function because it is constructed out of the warp-factor ${\rm H}(y)$, that appears in the metric ansatze \eqref{iibmet} or \eqref{mup}, as well as the non-locality function $\mathbb{F}^{(1)}(x - z)$ which is defined using gaussian functions (see \cite{desitter2}). 
Putting everything together, the contribution to the EOM \eqref{evaB1111} from \eqref{sarbutler}, appears to be:

{\footnotesize
\bg\label{ISOYG}
{\delta {\bf S}_{\rm inst}\over \delta {\bf C}_{012}(z, g_s)} = \lambda {\partial \over \partial z^{\rm M}} 
\left(\sqrt{g_6(z, g_s)}~\mathbb{F}(z) \mathbb{V}_2(z, g_s)  \mathbb{Y}_4^{(\{l_i, n_i\})012{\rm M}}(z)\right) 
\delta\left(\theta_k - {4\over 3}\right) + {\cal O}(g_s^{2/3}), \nd}
where we have assumed that $\lambda \mathbb{F}(z)$ remains finite with
$\lambda  = \sum_{q\ge 1} q c_q$.  The above equation tells us that there {\it is} a time-independent contribution to \eqref{evaB1111}, but it appears only as a {\it total derivative}. This means it doesn't contribute globally, {\it i.e.} it doesn't contribute to \eqref{palstodio}. This is unfortunate because if it cannot contribute globally, it cannot change our earlier conclusion of the absence of time-independent parts of the G-flux components ${\bf G}_{{\rm MNP}a}$.  Note that the total-derivative behavior that we saw above is generic: it will appear in the same way to the other G-flux EOM, namely \eqref{evaooo7}, and therefore cannot contribute to \eqref{shakalmey}. Any other non-perturbative effects also {\it cannot} contribute because:
\bg\label{9thgate}
{\rm exp}\left(-{1\over g_s^{2n/3}}\right) \to 0, \nd
for $n \in {\mathbb{Z}\over 2}$ and $g_s << 1$. Thus it seems that, even if we incorporate non-perturbative or non-local interactions, it is hard to get time-independent components of the G-flux. This is the reason why all entries in 
{\bf Table \ref{firzas2}} have $k \ge 1$, and it matches very well with our earlier conclusions in \cite{desitter2, desitter3, coherbeta} where we took $k \ge {3\over 2}$ for all the G-flux components. The fact that such choice of G-fluxes can still stabilize the moduli is explained in \cite{coherbeta} wherein we termed such moduli-stabilization as the {\it dynamical moduli stabilization}. 

\subsection{Superpotential, quantum corrections and time-dependences \label{superpo}}

The reason for considering the specific G-flux components ${\bf G}_{{\rm MNP}a}$ is because it dualizes to the three-form fluxes ${\bf H}_3$ and ${\bf F}_3$ in the IIB side via the following familiar decomposition:
\bg\label{soledadm}
{\bf G}_{{\rm MNP}a}~dy^{\rm M} \wedge dy^{\rm N} \wedge dy^{\rm P} \wedge dy^{a} = 
{\bf H}_3 \wedge dx^3 + {\bf F}_3 \wedge dx^{11}, \nd
which will tell us that once ${\bf G}_{{\rm MNP}a}$ become $g_s$ dependent functions, the corresponding IIB three-forms should also become $g_s$ dependent. There are also other components of the G-flux, defined over the eight-manifold, that we classified in {\bf Table \ref{firzas2}} which would contribute to the GVW 
\cite{GVW} superpotential. To see the precise behavior of the superpotential, let us define the $g_s$ behavior of the various four-forms entering the GVW superpotential as:

{\footnotesize
\bg\label{linroma}
&& \Omega_4(y, g_s) \equiv \sum_{k \ge 0} \Omega_4^{(k)}({y})\left({g_s \over {\rm H}}\right)^{2k/3}
\nonumber\\
&&{\bf G}_4(y, g_s) = {\bf G}_{\rm N_1 N_2 N_3 N_4}(y, g_s)~dy^{\rm N_1} \wedge dy^{\rm N_2} \wedge dy^{\rm N_3}
 \wedge dy^{\rm N_4} = \sum_{k \ge 1} {\bf G}_4^{(k)}({y}) \left({g_s\over {\rm H}}\right)^{2k/3}, \nd}
 where keeping up with the discussion above we take $k \ge 1$ for the $g_s$ expansion of the G-flux components; whereas $k \ge 0$ remains the $g_s$ expansion for the four-form $\Omega_4$. It is of course possible that $\Omega^{(0)}_4(y)$ vanish, but we leave the possibility that it could be non-zero, but 
 ${\bf G}_4^{(0)} = 0$. Using \eqref{linroma},  the GVW superpotential takes the following 
 form\footnote{We use ${\bf W}$ to denote the superpotential from M-theory, whereas $W$ is used to denote the superpotential in type IIB theory.}:

{\footnotesize
\bg\label{meltherry}
{\bf W} = \int {\bf G}_4 \wedge \Omega_4 = \sum_{k_i \in {\mathbb{Z}\over 2}}\int {\bf G}_4^{(k_1)} \wedge \Omega_4^{(k_2)} \left({g_s\over {\rm H}}\right)^{2(k_1 + k_2)/3} \equiv \sum_{k \in {\mathbb{Z}\over 2}}{\bf W}^{(k)}
\left({g_s\over {\rm H}}\right)^{2k/3}, \nd}      
where the integral is over the usual eight-manifold internal space. The $g_s$ dependence of the superpotential makes the story more non-trivial but is a necessary consequence of the underlying temporal dependences of the degrees of freedom. In the usual time-independent compactification of the form 
\cite{GVW, DRS} the superpotential is used to stabilize the complex structure moduli\footnote{This also translates to the background of the form \eqref{karintag} or \eqref{preacher}. The three-form fluxes supporting such background remains time-independent, implying that the M-theory uplifts of these backgrounds allow time-independent G-fluxes.}. 
Here neither the eight-manifold, nor the six-dimensional base are necessarily complex manifolds. In addition to that, they are also non-K\"ahler and time-{\it dependent} manifolds. So stabilization of the complex and the K\"ahler structure moduli are bound to become more intricate here and involve new procedures which we termed as the 
{\it dynamical moduli stabilization} in \cite{coherbeta}. 

We will not discuss the dynamical moduli stabilization here, as the analysis is more appropriately presented 
when de Sitter space is viewed as a Glauber-Sudarshan state \cite{coherbeta}. When the de Sitter space is viewed as a 
{\it vacuum}, which is what we take here, the moduli stabilization is much more involved and doesn't have the elegant flair attributed to it being a {\it state}. Nevertheless, some aspects of the superpotential can be studied at this stage. First however it is easy to see that the zeroth order superpotential always vanish, 
{\it i.e.}:
\bg\label{ivylabi}
{\bf W}^{(0)} = 0, \nd
simply because of the absence of time-independent G-flux components as elaborated in the earlier subsections. In the time-independent supersymmetric case, namely the one associated with the 
${\rm AdS}_4$ background 
\eqref{preacher}, the superpotential vanishes to restore the underlying supersymmetry. Clearly going from the M-theory uplift of \eqref{preacher}, which is supersymmetric, to \eqref{mup}, which is non-supersymmetric, the G-flux components entering the superpotential go from time-independent to time-dependent ones. Can we keep ${\bf W}^{(k)} = 0$ for $k \ge 1$? This would be possible for very special choices of the G-flux components ${\cal G}^{(k)}_{\rm N_1N_2N_3N_4}$ with ${\rm N_i} \in {\cal M}_4 
\times {\cal M}_2 \times {\mathbb{T}^2\over {\cal G}}$ in \eqref{kirdunst}. 

Even if we ignore for the time-being the dynamical stabilization of the complex structure moduli, there are many questions that can arise from such a choice of the superpotential. Foremost of them is whether we can maintain this condition once we switch on fermionic condensates on the seven-branes. A formal analysis of this will require us to compare two scenarios: one, in the absence and the other, in the presence of the fermionic condensates. In either cases, {\it i.e.} in the {\it absence} or {\it presence} of the condensates, the most efficient way to study this is from the background EOMs. These are the Einstein's equations discussed in much details in \cite{desitter2}, and for us it takes the following form: 
\bg\label{robbieM}
{\bf R}_{\mu\nu} - {1\over 2} {\bf g}_{\mu\nu} {\bf R} = {\bf T}^{({\rm flux})}_{\mu\nu} + {\bf T}^{({\rm b})}_{\mu\nu} + 
{\bf T}^{({\rm p})}_{\mu\nu} + {\bf T}^{({\rm np})}_{\mu\nu}, \nd
where the superscripts $\left({\rm flux, b, p, np}\right)$ denote the fluxes, branes, perturbative and non-perturbative 
respectively. The perturbative corrections come from the quantum series in \eqref{phingsha3} whereas the non-perturbative ones come from the last entry in the action \eqref{sheela2}. The fermionic condensate terms can come from the IIB seven-branes or from the bulk, and we will discuss them separately soon.
First however we should look at the $g_s$ scaling of each of the terms in \eqref{robbieM}. The LHS of 
\eqref{robbieM}, {\it i.e.} the Einstein tensor scales as \cite{desitter2}:
\bg\label{kidtheron}
{\bf R}_{\mu\nu} - {1\over 2} {\bf g}_{\mu\nu} {\bf R} = {1\over g_s^2} \left(3\Lambda 
+ {{\rm R} \over 2{\rm H}^4} - {4(\partial{\rm H})^2 \over {\rm H}^6}
+ {\square{\rm H}^4 \over 2 {\rm H}^8} + {\cal O}(g_s^{2k/3}\right)\eta_{\mu\nu}, \nd
where we used the background \eqref{mup} to get the RHS of \eqref{kidtheron} and ${\rm R}$ is the $g_s$ independent part of the eleven-dimensional curvature scalar. The additional $g_s^{2k/3}$ for $k \in {\mathbb{Z}\over 2}$ come from the ${\rm F}_i(g_s)$ factors in \eqref{mup}.  The cosmological constant 
$\Lambda$ is not a quantity added by hand here, rather it also appears from the metric ansatze \eqref{mup}. 
The question that we want to ask is: what are the perturbative and non-perturbative corrections that we need to add to order $g_s^{-2}$? The energy-momentum tensors for the G-flux components as well as the M2-branes do come proportional to $g_s^{-2}$, so we won't worry about them right now.  

The most dominant non-perturbative contributions come from the BBS -type  \cite{BBS} instanton gases as shown in \cite{coherbeta}. The other non-local instanton gases, like delocalized BBS, KKLT \cite{KKLT} or even delocalized KKLT type instanton gases, where the {\it delocalization} pertains to how the warped volume $\mathbb{V}_2$ in \eqref{sheela2} is considered (in one case it is related to the volume of 
fiber torus ${\mathbb{T}^2\over {\cal G}}$ and the other case it is related to the volume of ${\cal M}_2$). The $g_s$ scaling of the BBS-type instanton gases is given from \eqref{ddeluca} and \eqref{phingsha3} by:
\bg\label{ddeluca2}
&& {2} \sum_{i = 1}^{27} l_i + \sum_{i = 0}^2 n_i  + l_{34} + 
l_{35} + {2}\left(k'_1 + 2\right)\left(l_{28} + l_{29} + l_{31}\right) \nonumber\\
&+& \left(2k'_2 + 1\right)\left(l_{30} + l_{32} + l_{33}\right) + {2}\left(k'_3 - 1\right)\left(l_{36} +
l_{37} + l_{38}\right) = 8,  \nd
implying that all {\it eight derivative terms}, including {\it all} quartic order curvature terms and eighth order G-flux terms, should be democratically considered. We {\it cannot} ignore any of these terms because they all scale in the same way\footnote{At least with respect to $g_s$. The scaling with respect to ${\rm M}_p$ is complicated by the fact that there could be derivatives acting along the ${\cal M}_2$ directions in 
\eqref{phingsha3}. In the presence of localized G-flux components ${\bf G}_{{\rm MN}ab}$, the derivative action can give rise to extra ${\rm M}_p$ factors changing the naive ${\rm M}_p$ scalings of each of the quantum terms. This is one of the reasons for taking {\it all} terms classified by \eqref{ddeluca2} when solving for the background EOM \eqref{robbieM}.}!  The readers may notice the similarity between what we say here and what is been discussed in \cite{Sethi:2017phn} regarding the quartic curvature terms. Our analysis, in some sense, goes beyond \cite{Sethi:2017phn} by emphasizing the importance of including all polynomial classified by \eqref{ddeluca2}. Taking these into account, 
the explicit expression for the energy-momentum tensor for the BBS-type instanton gas may be written as the following integral form:

{\footnotesize
\bg\label{theron2}
{\bf T}_{\mu\nu}^{\rm BBS}(z, g_s) &= & 
-2\sum_{\{l_i\}, n_i}\int d^{11} x \sum_{k \ge 1}  c_k~
{\rm exp}\Big[- k {\rm M}_p^6 \int d^6 y \sqrt{{\bf g}_6(y, g_s)}
~\mathbb{F}^{(1)}(x - y) \mathbb{Q}_{\rm T}^{(\{l_i\}, n_i)}(y, g_s)\Big]\nonumber\\
&&~~~~~~\times ~ \mathbb{F}^{(1)}(x - z)\sqrt{{\bf g}_6(z, g_s)} 
~{\delta \mathbb{Q}_{\rm T}^{(\{l_i\}, n_i)}(z, g_s) \over \delta {\bf g}^{\mu\nu}(z, g_s)}  
\sqrt{{\bf g}_{11}({\bf x}, g_s) \over {\bf g}_{11}(z, g_s)}, \nd}
where $\mathbb{Q}_{\rm T}^{(\{l_i\}, n_i)}(z, g_s)$ contains {\it all} the terms classified by \eqref{ddeluca2}, and note the absence of the $\mathbb{V}_2(y, g_s)$ factor. This will appear for the delocalized BBS-type 
instanton gases which are classified by $\theta_k \equiv \theta_{lkk'\hat{k}} = {4\over 3}$ in \eqref{ddeluca}. As mentioned above, all these have to be taken together. In addition to that there are also perturbative contributions coming directly from \eqref{phingsha3}. They can be easily classified, see \cite{desitter2}, so we will not elaborate them here. 

The integral along $x$ in \eqref{theron2} can be shown to be finite. This is more or less similar to the finiteness of \eqref{ckeaton} as the exponential terms in \eqref{theron2} scale as $g_s^{2/3}$ so have perturbative expansions (provided we take the simplest case where $\mathbb{F}^{(1)}(x-z)$ does not have any $g_s$ scaling). The sum over $k$ in \eqref{theron2} is convergent because for large $k$, the exponential term dies off (in addition to that we can have a perturbative expansion of the exponential factor when $g_s < 1$). This convergence is important, but one might worry about the volume term in the first line of \eqref{sheela2} as it comes with sum over all $c_k$. This might seem to blow up, but a careful study of the energy-momentum tensor in \eqref{theron2} reveals another equivalent term that exactly cancels the 
troublesome factor thus retaining only the expression showed in \eqref{theron2} without any extra pieces. 

The question that we want to ask is how \eqref{robbieM} changes once fermionic condensates from the type IIB seven-branes are taken into account. As mentioned earlier, these seven-branes dualize to seven (not eight!) -dimensional surfaces in M-theory. The energy-momentum tensor from the fermionic condensates on a given seven-brane takes the following form:

{\footnotesize
\bg\label{mutachat3} 
{\bf T}^{({\rm ferm})}_{\mu\nu}(\bar{z}) = - \sum_{i, q} {2{\bf T}_7  \over \sqrt{-{\bf g}_{11}(\bar{z})}} 
\int {d^7 y \left(\bar{\bf\Psi}{\bf\Psi}\right)^q  \over  
{\rm M}_p^{\sigma^{(i)} - 7}}\left[ 
\sqrt{-{\bf g}_7(\bar{y})} ~
{\delta \over \delta {\bf g}^{\mu\nu}(\bar{z})} +
{\delta \left(\sqrt{-{\bf g}_7(\bar{y})}\right) \over \delta {\bf g}^{\mu\nu}(\bar{z})} 
\right]{\widetilde{\mathbb{Q}}}_{\rm T}^{(i, r)} (\bar{y}), \nd} 
where we used compact notations $\bar{z}$ and $\bar{y}$ to denote $(z, g_s)$ and $(y, g_s)$ respectively, and ${\widetilde{\mathbb{Q}}}_{\rm T}^{(i, r)} (\bar{y})$ is defined in the fourth line of \eqref{sheela2}. The 
$g_s$ scaling of \eqref{mutachat3} is easy to work out and is given by:
\bg\label{sessence}
n_1 + n_2 + 2\sum_{i = 1}^{27} l_i + q + 3 \sum_{r = 0}^2l_{36 + r} = 4, \nd
with the parameters as defined in \eqref{phingsha3} and \eqref{ddeluca}. We can easily see that there are multiple solutions to \eqref{sessence} and, as before, all of these have to be taken together to study the contributions of the fermionic condensates on the EOM \eqref{robbieM}. 

Unfortunately these are not the only terms with fermionic condensates. One may even replace the G-flux components entering $\mathbb{Q}_{\rm T}^{(\{l_i\}, n_i)}(y, g_s)$ in \eqref{phingsha3} by fermionic extensions somewhat along the lines of \eqref{emmarob}, but now involving all other internal G-flux components (even the curvature terms could be given a similar treatment, much like the fermionic extensions of all the bosonic degrees of freedom dealt in \cite{fermions}). Such quantum series could replace $\mathbb{Q}_{\rm T}^{(\{l_i\}, n_i)}(y, g_s)$ in \eqref{theron2}, and are therefore classified by 
\eqref{ddeluca2}. Clearly {\it all} of these are needed if we want to make sense of \eqref{robbieM}, but 
despite this proliferation of the quantum terms, they are still {\it finite} in number (albeit large). 

Let us now answer the question that we posed earlier: how do the addition of the fermionic condensates effect the EOM \eqref{robbieM}? Clearly the LHS of \eqref{robbieM}, {\it i.e.} \eqref{kidtheron} cannot change because the background is fixed to \eqref{mup}. We would also like to fix the G-flux components such that they continue to satisfy ${\bf W}^{(k)} = 0$ for $k \ge 1$. This way the energy-momentum contributions from G-fluxes, namely ${\bf T}^{({\rm flux})}_{\mu\nu}$, remain unchanged. The perturbative terms, {\it i.e.} 
${\bf T}^{({\rm p})}_{\mu\nu}$, only contribute as $\theta_{k} \equiv \theta_{lkk'\hat{k}} = {2\over 3}$ so they can only renormalize the flux factors from ${\bf T}^{({\rm flux})}_{\mu\nu}$. This means the only possible way to compensate for the addition of fermionic condensates is to impose the following consistency condition:
\bg\label{ledgeman}
{\bf T}^{({\rm ferm})}_{\mu\nu}(z, g_s) = {\bf T}^{({\rm np})}_{\mu\nu}(z, g_s)  - 
\widetilde{\bf T}^{({\rm np})}_{\mu\nu}(z, g_s), \nd
where we have ignored the M2 and $\overline{\rm M2}$ branes' contributions because when they are kept well separated they contribute equally in the presence or absence of the fermionic condensates (we are ignoring spacetime fermionic interactions, and the possible {\it fractional} nature of these branes). The equation \eqref{ledgeman} is interesting in many sense. It tells us that once we switch on fermionic condensates, keeping the G-flux components unchanged, we will have to change the coefficients of the curvature terms that enter the non-perturbative energy-momentum tensor. Thus $\widetilde{\bf T}^{({\rm np})}_{\mu\nu}(z, g_s)$ and ${\bf T}^{({\rm np})}_{\mu\nu}(z, g_s)$ differ in how the quantum terms from 
\eqref{phingsha3} are arranged. If we ignore the fermionic extension of 
$\mathbb{Q}_{\rm T}^{(\{l_i\}, n_i)}(y, g_s)$, then the LHS of \eqref{ledgeman} is classified by 
$\theta_{k} \equiv \theta_{lkk'\hat{k}} = {4\over 3}$ in \eqref{ddeluca}. On the other hand, the RHS of 
\eqref{ledgeman} gets contributions from four sources: BBS, delocalized BBS, KKLT and delocalized KKLT 
-type instanton gases. The BBS instanton gases, as shown in \eqref{theron2}, scale as 
$\theta_{k} \equiv \theta_{lkk'\hat{k}} = {8\over 3}$, whereas the other three scale as 
$\theta_{k} = {4\over 3}, {2\over 3}$ and ${4\over 3}$ respectively (see sections 3.2 and 3.3 of \cite{coherbeta}). 
The delocalized BBS  is defined by incorporating the volume of the fiber torus ${\mathbb{T}^2\over {\cal G}}$
in the expression \eqref{theron2}, whereas the delocalized KKLT is defined by incorporating the volume of 
${\cal M}_2$ in an expression similar to \eqref{theron2}. Since they both scale as $\theta_k = {4\over 3}$, it appears that \eqref{ledgeman} can be solved by incorporating the effects of these two instanton gases. 

Such an analysis then explains what effects do fermionic condensates typically have. Once we switch on fermionic condensates the quadratic curvature terms as well as the quartic G-flux interactions change. This change {\it absorbs} the backreaction effects of the fermionic condensates here. Of course this is the simplest situation, because we have ignored many things. However a quantitative analysis like 
\eqref{ledgeman} at least suggests that a similar analysis in the presence of more complicated fermionic interactions may be easily carried out.

\section{Discussions and conclusions}\label{sec5}
The motivation for this work was to construct a $4$-dimensional de Sitter space within string theory which is compatible with the so-called swampland conjectures. In particular, our motivation was to look for a dS solution whose lifetime is short-enough to be consistent with the TCC. Since it is well-known that the standard KKLT scenario (as well as the LVS construction) has lifetimes which far exceed the bound set by the TCC, our aim was to incorporate other stringy ingredients such that we end up with short-lived dS vacua obeying the TCC bound. Note that in no way are we claiming to show evidence for the TCC or the lack of it. We simply assume the bounds implied by the TCC on the lifetime on \textit{any} dS vacua and try to build one such solution within string theory.

At first sight, it seems impossible to achieve such an objective since the lifetime associated, say, with the KKLT formalism, is fixed by the gravitational Coleman-de Luccia instanton to the asymptotic Minkowski volume at infinite volume. And this universal tunneling rate is shown to be of the same order of magnitude for all such dS solutions \cite{Westphal:2007xd}, and is unacceptably large compared to the TCC bound. However, this result also provides us a glimpse into the possible solution -- to have more than one AdS vacua prior to uplifting. In our first pass, we showed that it is indeed possible to build such a potential having multiple AdS vacua prior to uplifting. However, instead of using an existing model, we improved upon the KL model to have only non-perturbative contributions coming from gaugino condensations to the superpotential. The reason for this choice was also to negate criticisms of \cite{Sethi:2017phn} and avoid having an inconsistent starting point when one considers a $W_0\neq 0$. Without going into the details of the derivation of such a solution from the $10$-d equations of motion, we find that indeed these dS vacua, made out of purely non-perturbative ingredients, have a lifetime many orders of magnitude smaller than the standard KKLT scenario and are, therefore, phenomenologically very different. The main reason for this is that there are now other channels for the dS vacuum to decay into as opposed to only having the supersymmetric Minkowski one at large volumes. Furthermore, this model has two maxima, after the uplift, which separates it from the KL model and is one of its distinctive features. Interestingly, this type of solutions might turn out to be useful when constructing inflationary backgrounds and this shall be further investigated in the future.

Unfortunately, however, we also find that the lifetimes of these dS vacua, short as they are, are still out of bounds of the TCC. The phenomenology of the solutions do not allow for the lifetimes to be shortened any further within the supergravity approximation assumed during their construction. Thus, although we manage to find new, and phenomenologically interesting, dS solutions in string theory, assuming the $4$-d effective field theory description, they still do not satisfy the bounds on the lifetime postulated by the TCC. Of course, it is worth mentioning here that if the TCC turns out to be an approximate condition, with possible refinements of it showing up \cite{Berera:2020dvn}, our solutions might turn out to be compatible with it after all.

Having constructed these short-lived dS vacua, next we wanted to reexamine some of our assumptions in the model-building from a higher-dimensional string theory perspective. In particular, we stated a set of four assumptions at the beginning of section \ref{sec4} to try and see if these can be self-consistently satisfied. Firstly, our analysis here has revealed  that we cannot keep the G-flux components time-independent implying that, in the dual IIB side, the three-form fluxes ${\bf H}_3$ and ${\bf F}_3$ cannot remain time-independent. Such a conclusion also extends to the superpotential and the fermionic condensates too: they appear to retain their temporal dependences to avoid inconsistencies. On the other hand, the condition $W_0 = 0$, which dualizes to ${\bf W}^{(k)} = 0$ for $k \ge 1$ appears to be possible, provided we have a consistency condition of the form \eqref{ledgeman}. Unfortunately, however, consistency of the background EOMs, force us to consider {\it all} quantum terms classified by $\theta_{lkk'\hat{k}} = {8\over 3}$ and ${4\over 3}$ in \eqref{ddeluca}. This shows that the suspicions raised in \cite{Sethi:2017phn} are realized in our work in a different way where the higher curvature terms make it necessary to turn the system time-dependent, and this is realized upon analyzing the uplifting procedure from the M-theory perspective. 

\subsection{de Sitter space: Vacuum or a Glauber-Sudarshan state?}
Our analysis reveals some of the technical complications as to why it is important to consider time-dependent fluxes as well as to include quantum terms of all orders to support a dS vacua. This brings us to our final question -- Can one overcome these impediments to realize dS spacetime in string theory? And the answer is a resounding yes, \textit{provided one views dS not as a vacuum but as a coherent state in the full theory}. This has been worked out in detail in \cite{coherbeta}, showing how one can successfully construct such a state in string theory, taking into account all the necessary local, nonlocal, non-perturbative and topological quantum corrections\footnote{The {\it first} realization of four-dimensional dS space as a coherent state appeared in \cite{dvali} (see also \cite{guisti}). The construction of \cite{dvali} did not involve any stringy ingredients. In \cite{coherbeta} we argued how such a construction may be realized in full string theory as a quantum mechanically {\it stable} configuration.}. Even more, using the Schwinger-Dyson equations, it was demonstrated that this state remains quantum-mechanically stable in the presence of this infinite set of quantum corrections coming from the full interacting M-theory action. This Glauber-Sudarshan state itself is constructed from mode fluctuations of the metric and G-fluxes on top of a solitonic, Minkowski, time-independent background with no running moduli forms. However, the crucial input for constructing the coherent state is that it is built upon the interacting vacuum of the full M-theory action, and not the usual harmonic free vacuum. In this setup, it was shown that the expectation values of the metric describe a dS spacetime, with time-dependent internal dimensions, invoking a  \textit{dynamical moduli stabilization} procedure to avoid the usual Dine-Seiberg runaway. The Glauber-Sudarshan state spontaneously breaks supersymmetry at scale that is much higher than that of the cosmological constant. The hierarchy between these two scales is due to the fact that the cosmological constant is suppressed by the unwarped volume of the $\mathcal{M}_4 \times \mathcal{M}_2$ internal manifold. 

Interestingly, the time for which the dynamics of the Galuber-Sudarshan state remains weakly coupled sets an upper bound on the lifetime of the dS solution, and is given by $\tau < 1/\sqrt{\Lambda}$, $\Lambda$ being the cosmological constant. Phenomenologically, this lifetime is completely consistent with the upper bound coming from the TCC although the real reason why such trans-Planckian censorship never plays any destabilizing role for such a coherent state description of dS has been detailed in \cite{coherbeta}. In a nutshell, the fluctuations on top of a dS background, in this formalism, are written as deformation of the Glauber-Sudarshan state and the time-dependent frequencies of the modes can be related to mode-fluctuations over the Minkowski background. This also provides an immediate understanding of the usual Gibbons-Hawking entropy of dS space as a microscopic entanglement entropy of the same fluctuations which build up the coherent state. Other instabilities, associated with the choice of the vacuum for quantum field theory on dS space \cite{Danielsson:2018qpa}, are also naturally avoided in such a coherent state picture. Therefore, it seems that considering a Glauber-Sudarshan state, one can not only find a meta-stable description of dS space in string theory, but also one that is compatible with the restrictions coming from the swampland. It would be interesting to study the phenomenological model studied in this paper with multiple AdS vacua from such a coherent state perspective in M-theory.

\subsection{Non-perturbative backgrounds and $\alpha'$-Cosmology}

In section \ref{sec4} we saw that having an hierarchical structure among the higher derivative terms constrains the time-dependence of the fluxes (see the discussion after equation (\ref{ddeluca})). As a result, as discussed in \cite{desitter2, desitter3}, having a dS background with time-independent fluxes require us to consider the entire tower of high derivatives corrections, signaling a break in the effective field theory description. That is, all the terms in the $g_s$ and ${\rm M}_p$ (equivalently $\alpha'$) expansions would need to contribute to the same order in the equations of motion. Recently, a similar condition appeared as crucial for the duality invariant setup of $\textit{$\alpha'$-Cosmology}$ \cite{Hohm:2019jgu,Bernardo:2019bkz}. Compared to the M-theory analysis, this framework involves only ${\bf H}_3$, backgrounds with $d$ Abelian isometries and includes all duality invariant higher order $\alpha'$ corrections to the action. Since the dilaton also takes part in the duality transformation, it may generally have time dependence. 

As in the M-theory analysis of section \ref{sec4}, solutions to the $\alpha'$-Cosmology equations of motion can be perturbative, \textit{i.e.}, constructed from solutions to the lowest order equations that receive corrections at each order in the $\alpha'$ expansion, or non-perturbative ones, requiring all corrections to exist. In fact, as shown in \cite{Hohm:2019ccp, Hohm:2019jgu}, the entire tower of corrections is necessary for having dS solutions in the string frame (with zero fluxes and a time-dependent dilaton). After including matter in the form of a perfect fluid energy-momentum tensor in the framework \cite{Bernardo:2019bkz}, solutions to the equations of motion were studied in \cite{Bernardo:2020zlc} and a model sourced by a string gas was introduced in \cite{Bernardo:2020nol}. This model has a short TCC compatible phase of accelerated (super-exponential) expansion that last long enough to solve the horizon problem of standard cosmology \cite{Bernardo:2020bpa}, depicting the phenomenological relevance of this class of backgrounds. 

The non-perturbative solutions mentioned above illustrate the fact that breaking the hierarchy of corrections in an expansion may not prevent us from having genuine backgrounds as non-perturbative solutions of the equations of motion. This goes in line with the idea that there can be dS backgrounds in string theory, but it could be that they are not accessible in the effective field theory approach, in which one has to truncate higher order corrections up to some finite order. A proper connection between $\alpha'$-Cosmology and the M-theory should be constructed to study these issues further.

\section*{Acknowledgments}
HB and SB would like to thank Evan McDonough for discussions during early stages of this work. SB thanks Daniel Junghans for interesting discussions.

\noindent HB and KD are supported in part by funds from NSERC and  from the Canada Research Chair program. SB is partially supported by a McGill Space Institute fellowship and a CITA National Fellowship.

\end{document}